\documentclass{aa}

\usepackage{graphicx}
\usepackage{txfonts}
\usepackage{subcaption}
\usepackage{gensymb}

\begin{document}

    \title{Measuring the orbit shrinkage rate of hot Jupiters due to tides}

    \author{
    N. M. Rosário\inst{\ref{inst1}}\fnmsep\inst{\ref{inst2}}
    \and
    S. C. C. Barros
    \inst{\ref{inst1}}\fnmsep\inst{\ref{inst2}}
    \and
    O. D. S. Demangeon
    \inst{\ref{inst1}}\fnmsep\inst{\ref{inst2}}
    \and
    N. C. Santos
    \inst{\ref{inst1}}\fnmsep\inst{\ref{inst2}}
    }

    \institute{
    Instituto de Astrofísica e Ciências do Espaço, CAUP, Rua das Estrelas, PT4150-762 Porto, Portugal\\ \email{Nuno.Rosario@astro.up.pt}\label{inst1}
    \and
    Departamento de Fisica e Astronomia, Faculdade de Ciências, Universidade do Porto, Rua do Campo Alegre 687, PT4169-007 Porto, Portugal\label{inst2}
    }

   \date{Received July 15, 2022 / Accepted October 20, 2022}
   
   % \abstract{}{}{}{}{} 
% 5 {} token are mandatory
 
  \abstract
   {A tidal interaction between a star and a close-in exoplanet leads to shrinkage of the planetary orbit and eventual tidal disruption of the planet. Measuring the shrinkage of the orbits will allow for the tidal quality parameter of the star ($Q'_\star$) to be measured, which is an important parameter to obtain information about stellar interiors.}
   {We analyse data from the Transiting Exoplanet Survey Satellite (TESS) for two targets known to host close-in hot Jupiters, which have significant data available and are expected to have a fast decay: WASP-18 and WASP-19. We aim to measure the current limits on orbital period variation and provide new constrains on $Q'_\star$ for our targets.}
   {We modelled the transit shape using all the available TESS observations and fitted the individual transit times of each transit. We used previously published transit times together with our results to fit two models, a constant period model, and a quadratic orbital decay model, using Markov chain Monte Carlo (MCMC) algorithms.}
   {We obtain new constrains on $Q'_\star$ for both targets and improve the precision of the known planet parameters with the newest observations from TESS. We find period change rates of $(-0.11\pm0.21)\times10^{-10}$ for WASP-18b and $(-0.35\pm0.22)\times10^{-10}$ for WASP-19b and we do not find significant evidence of orbital decay in these targets. We obtain new lower limits for $Q'_\star$ of $(1.42\pm0.34)\times10^7$ in WASP-18 and $(1.26\pm0.10)\times10^6$ in WASP-19, corresponding to upper limits of the orbital decay rate of $-0.45\times10^{-10}$ and $-0.71\times10^{-10}$, respectively, with a 95\% confidence level. We compare our results with other relevant targets for tidal decay studies.}
   {We find that the orbital decay rate in both WASP-18b and WASP-19b appears to be smaller than the measured orbital decay of WASP-12b. We show that the minimum value of $Q'_\star$ in WASP-18 is two orders of magnitude higher than that of WASP-12, while WASP-19 has a minimum value one order of magnitude higher, which is consistent with other similar targets. Further observations are required to constrain the orbital decay of WASP-18 and WASP-19.}
   

   \maketitle
%
% ==================================================

\section{Introduction}

In 1995, 51 Pegasi b became the first exoplanet discovered orbiting a solar-like star \citep{mayor1995jupiter}. It is a massive planet, about half the mass of Jupiter, with a short period, orbiting close to its host star. These planets, now known as hot Jupiters due to their size and high temperature, have been found around several systems and provide an astronomical laboratory for our physical models of planetary formation and evolution, as well as planet-star interaction.  Hot Jupiters are an uncommon presence around Sun-like stars when compared to smaller planets \citep{howard2012planet, wright2012frequency, johnson2010giant}, and the circumstances of their formation and evolution are still not completely understood \citep[e.g.][]{dawson2018origins}. Due to their size and short period, these planets have a high signal-to-noise ratio (S/N), making it easier to obtain data on them and turning hot Jupiters into an important source of information regarding the physics of planet formation and evolution.

Due to the proximity to their host star, hot Jupiters are strongly affected by tidal forces. In cases where the rotation of the star is not synchronised with the orbital period of the planet, a transfer of angular moment occurs, leading to an increase in the rotation of the star and a shrinkage of the planet's orbit. This effect, known as tidal decay \citep{counselman1973outcomes, levrard2009falling, matsumura2010tidal}, gives important information about the tidal interaction between a planet and star and helps to constrain stellar physics through the tidal quality factor $Q'_\star$. This parameter measures the efficiency of the energy dissipation during tidal interactions and is related to the propagation of tidal oscillations inside the star \citep{ogilvie2007tidal, ogilvie2014tidal}. According to the theory developed by \citet{zahn1975dynamical, zahn1977tidal}, larger stars with convective cores and radiative envelopes are expected to have a less efficient dissipation and, therefore, a larger value of $Q'_\star$.\par

Other physical processes can, in the short term, result in a shift in period that is similar to tidal decay: apsidal precession \citep{miralda2002orbital, heyl2007using, jordan2008observability, ragozzine2009probing} and the presence of external bodies affecting their orbits \citep{gibson2009transit, gibson2010transit}. Apsidal precession is the rotation of the ascending node along the orbital plane and as such, this effect is periodic and does not lead to a long-term decay of the orbit. Additionally, a change in period may arise from gravitational effects from nearby companions disturbing the orbit. An example of this effect is reported by \cite{triaud2017peculiar} in their study of the system architecture of WASP-53 and WASP-81. \cite{turner2021characterizing} have recently suggested that the orbit of WASP-4b \citep{wilson2008wasp} may also be changing because of another companion nearby.  \par

WASP-12b, a hot Jupiter orbiting a late F-type star, was the first target confirmed as having a decaying orbit caused by tidal interactions \citep{patra2017apparently, maciejewski2018planet, turner2021decaying, wong2022tess}. It orbits its star with a period of $1.09$ days, and has a mass of $1.5\ M_\mathrm{J}$ and a radius of $1.90\ R_\mathrm{J}$. Its orbit is shrinking at a rate of $29.81\pm0.94$ ms $\mathrm{yr^{-1}}$, leading to a decay timescale of $3.16\pm0.10$ Myr. The tidal quality factor of $Q'_\star=(1.50\pm0.11)\times10^5$ is thought to be on the lower side of the predicted values \citep{wong2022tess}. 

Despite being one of the prime targets to measure tidal decay \citep{patra2020continuing}, so far there is still no evidence towards a decaying period in WASP-18b, with \cite{maciejewski2020planet} placing a lower limit for $Q'_\star$ of $3.9\times10^6$ at 95\% confidence level. The value is of the same order of magnitude as the ones found by \cite{shporer2019tess} and \cite{patra2020continuing}, which also investigated this system.
WASP-19b \citep{hebb2009wasp}, which is one of the shortest period hot Jupiters reported to date (0.78 days), is also an interesting target for tidal decay studies. Several authors have studied this system \citep[e.g.][]{hellier2009orbital,tregloan2013transits, espinoza2019access, petrucci2020discarding} and \citet{patra2020continuing} found statistically significant evidence towards a decreasing period, with $\dot{P}=dP/dt=(-2.06\pm0.42)\times 10^{-10}$. However, WASP-19 is a highly active star, leading to distortions in the light curves thought to be caused by star spots \citep{tregloan2013transits, espinoza2019access}. The result by \cite{petrucci2020discarding}, discarding tidal decay in WASP-19, contradicts these previous results, finding a value of $\dot{P}=0.0114\pm0.74\times10^{-10}$ consistent with zero, and increases the importance of continuing observations of this system.\par

In this paper, we present our analysis of the Transiting Exoplanet Survey Satellite \citep[TESS;][]{ricker2014transiting} data concerning two hot Jupiters expected to show orbital decay, WASP-18b and WASP-19b. We aim to search for signs of tidal decay on these targets and to measure the current constrains on the tidal decay parameters, which allow us to better understand the migration mechanisms of hot Jupiters and the reasons behind the decaying orbits.\par
We describe our target systems and the process of target selection in Sect. 2. In Sect. 3, we present the TESS data and the pre-processing of the light curves. The transit analysis is listed in Sect. 4 and in Sect. 5 we describe our transit timing variation (TTV) and orbital decay models. We present and discuss the results of the TTV analysis in Sect. 6 and provide a short summary of the current status of orbital decay studies in other systems in Sect. 7. Finally, a summary and conclusion is presented in Sect. 8.

% =================================================================================
% SECTION 2 - TARGET SELECTION
% =================================================================================

\section{Target selection}

In order to work with the most  relevant targets we looked at several hot Jupiters that orbit close to their stars. We mainly focussed on the expected orbital decay rate of those planets in order to explore the most promising candidates. \citet{patra2020continuing} plotted the scaled orbital semi-major axis $a/R_\star$ against the planet-to-star mass ratio $M_\mathrm{p}/M_\star$, two quantities that are proportional to the decay rate according to the current orbital decay models, and they provided several cutoffs based on the decay rate for a constant $Q'_\star$ in a population of cold and hot stars brighter than $V=14.5$. We followed the approach taken in that paper and started by looking at a sample of 12 targets, six candidates orbiting hot stars ($T_\mathrm{eff} > 6000$ K) and six orbiting cold stars, with an expected decay rate of $dP/dT<10^7$ yrs for a $Q'_\star = 10^6$.\par

We then looked at the targets that have been observed by TESS in more than one season and whose data were not already published, so as to be able to obtain the maximum amount of transits and therefore obtain the best possible constrain on the orbital decay parameters. We also looked at the previous available data from other observations to ensure the data span a large enough timescale to possibly detect a change in period. For example, HATS-18 \citep{penev2016hats} would seem promising with the previous criteria; however, the lack of previous data makes it hard, if not impossible, to properly search for period changes. WASP-4 \citep{wilson2008wasp} and WASP-43 \citep{hellier2011wasp}, for example, would also make for good targets in this study; however, orbital decay analyses using the available data from TESS were recently published (WASP-4: \citealp{turner2021characterizing}; WASP-43: \citealp{davoudi2021investigation}).\par

According to \cite{patra2020continuing}, it is expected that planets around colder stars have a faster tidal dissipation due to their larger convective envelopes. Therefore, it can be important to investigate tidal decay around both hot and cold stars. Given the previous criteria and taking stellar temperature into account, we looked into the best candidates orbiting a hot star and the best candidates orbiting a cold star. From the original sample, WASP-18 is shown to be one of the most promising candidates for tidal decay thanks to its high $M_\mathrm{p}/M_\star$ ratio. Furthermore, it was observed by TESS in a total of four sectors, which gives us a good amount of data to work with, and it has been the subject of several decay studies and previous observations \citep[e.g.][]{shporer2019tess, maciejewski2020planet}. WASP-19 is one of the most promising targets among cold stars. It has available data from TESS in two sectors and has been studied for several years \citep[e.g.][]{mancini2013physical, espinoza2019access}, with observations from different instruments in both photometry and spectroscopy, in part due to the high stellar activity that is present in the star. The additional TESS data may also provide new input to understand the effect of the activity in the orbital decay studies.

% ==================================================

\subsection{WASP-18}

WASP-18b \citep{hellier2009orbital, southworth2009physical} is a massive hot Jupiter with a mass of 11.4 $M_\mathrm{J}$ and a radius of 1.20 $R_\mathrm{J}$, orbiting its star on a period of 0.94 days. The star, WASP-18, is a bright ($V=9.30$) F-type star with a mass of 1.46 $M_\sun$, an effective temperature of 6431 K, and an estimated age of 1 Gyr \citep{patra2020continuing}. Several authors have looked at this system before to try to find evidence of tidal decay since, according to theoretical models, WASP-18b should be one of the fastest decaying targets we know of, in part due to its large mass relative to the star. However, \cite{wilkins2017searching} initially found no evidence of rapid decay. \cite{mcdonald2018pre} included an early transit for \textit{Hipparcos} and found a weak (1$\sigma$) sign of decay, but one must take the high uncertainty in the timing of the \textit{Hipparcos} transit  into account. With the first TESS observations for WASP-18, in Sectors 2 and 3, new constrains have been found in the tidal decay parameters, with \cite{shporer2019tess} and \cite{maciejewski2020planet} reporting results consistent with the findings of \cite{wilkins2017searching}, with the new data not showing enough evidence to confirm the decay of the orbit of WASP-18b.
With two additional TESS sectors and a 2-year gap between observations, we expect to add an additional constraint to the tidal decay parameters and strengthen the evidence towards the presence or absence of rapid decay.

% ==================================================

\subsection{WASP-19}

WASP-19b \citep{hebb2009wasp} is a Jupiter-sized planet with 1.14 $M_\mathrm{J}$ and 1.4 $R_\mathrm{J}$ orbiting an active solar-type G-star with around 10 Gyr, a mass of 0.9 $M_\sun$, and an effective temperature of 5460 K. It is one of the shortest-period hot Jupiters discovered so far, with a period of 0.78 days, and it was surpassed recently by the discovery of TOI-2109 with a period of 0.67 days \citep{wong2021toi}. With the high amount of activity reported, WASP-19 is an interesting but challenging target to measure orbital decay. Several authors have obtained WASP-19b transits \citep[e.g.][]{mancini2013physical, espinoza2019access}, and \cite{patra2020continuing} added their own transits to previous data and found strong evidence of decay, with a period change rate of $-6.50\pm1.33$ ms yr$^{-1}$. However, they suspect stellar activity is causing systematic errors and the result may be misleading. This evidence is further weakened by \cite{petrucci2020discarding}, who include several more recent transits from the literature and from their own observations and show evidence towards a constant period, with a value of $\dot{P}=0.015\pm0.96$ ms yr$^{-1}$ consistent with zero. Despite this, they provide an upper limit for $\dot{P}$ of -2.294 ms yr$^{-1}$. The newest TESS observations might shed some new light on WASP-19b and the true nature of its orbital period and help to clarify the behaviour of this hot Jupiter.

% =================================================================================
% SECTION 3 - TESS DATA
% =================================================================================

\begin{figure}
     \centering
     \begin{subfigure}{\linewidth}
         \centering
         \includegraphics[width=\linewidth]{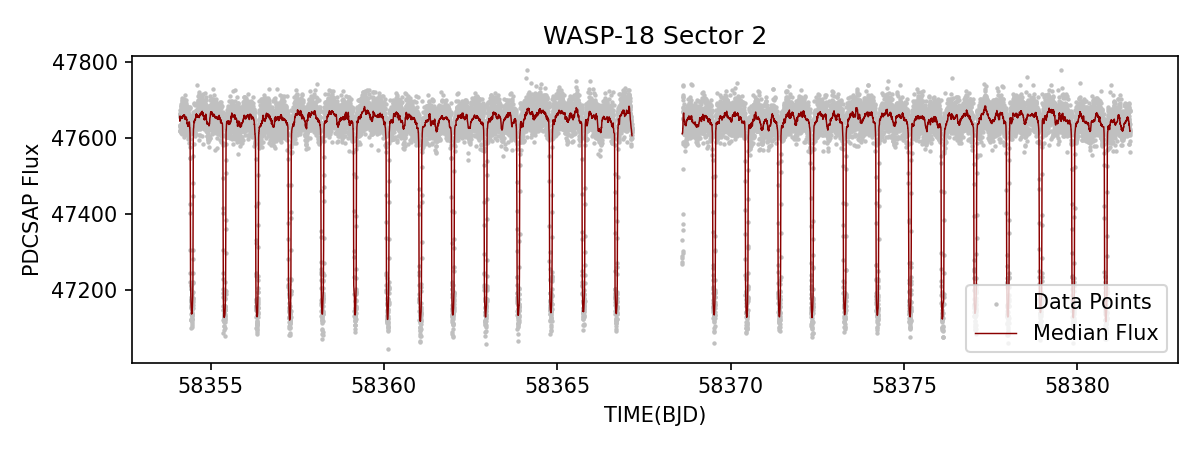}
         \label{wasp18_2}
     \end{subfigure}
     \begin{subfigure}{\linewidth}
         \centering
         \includegraphics[width=\linewidth]{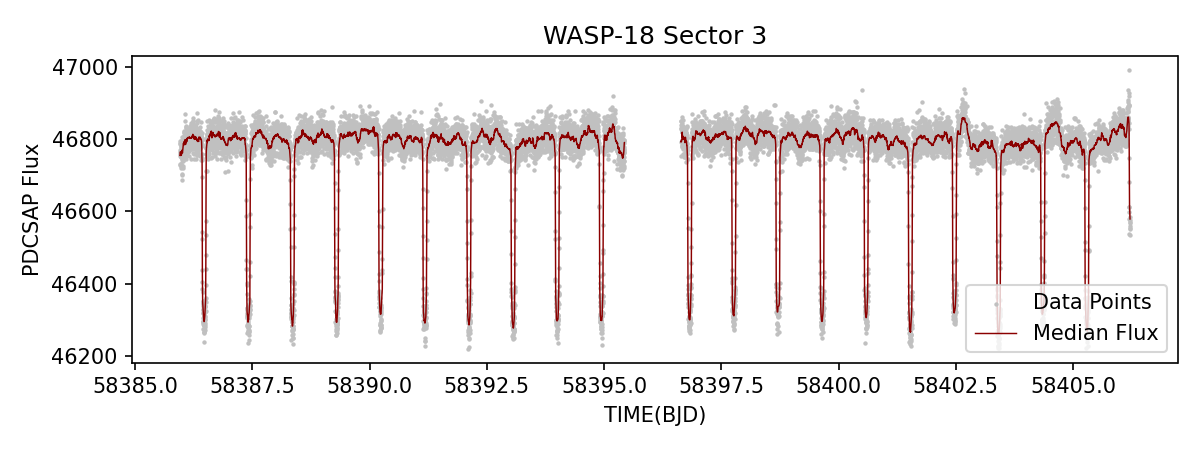}
         \label{wasp18_3}
     \end{subfigure}
     \begin{subfigure}{\linewidth}
         \centering
         \includegraphics[width=\linewidth]{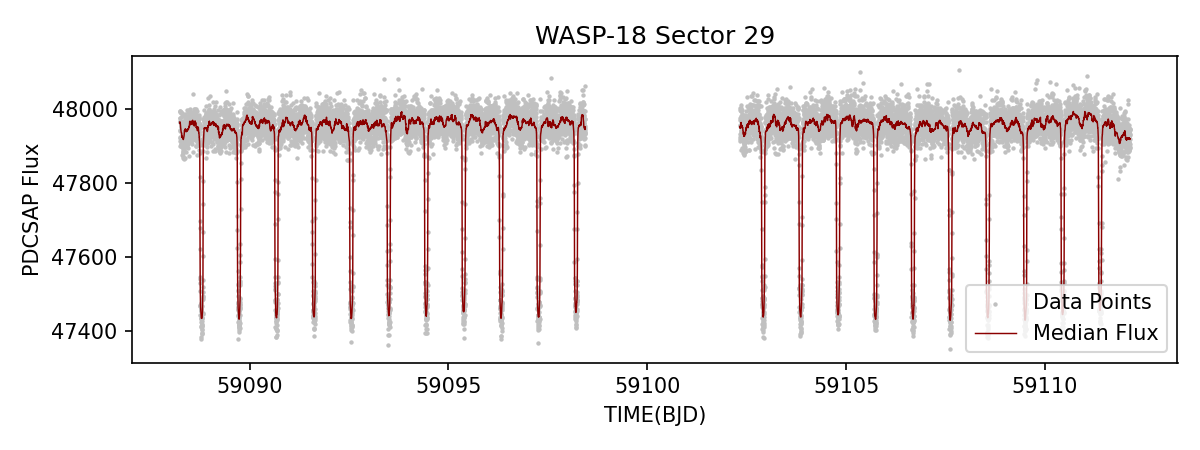}
         \label{wasp18_29}
     \end{subfigure}
     \begin{subfigure}{\linewidth}
         \centering
         \includegraphics[width=\linewidth]{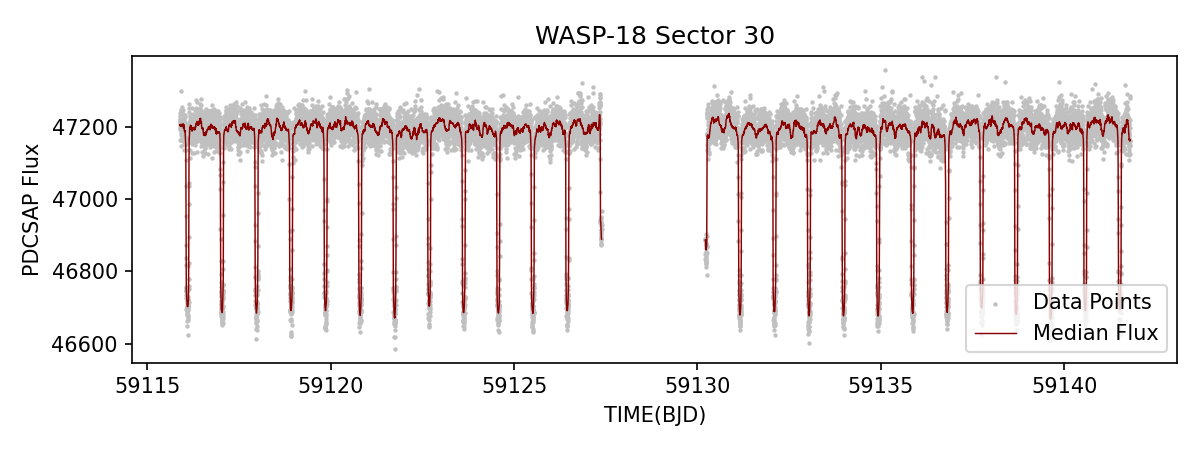}
         \label{wasp18_30}
     \end{subfigure}
        \caption{Plot of the PDC light curves of WASP-18 with NaNs and quality flagged points removed. Moving median filter over 32 points shown in red. Incomplete transits are shown near the gaps that separate the two physical orbits of the spacecraft.}
        \label{wasp18LC}
\end{figure}

\begin{figure}
     \centering
     \begin{subfigure}[b]{\linewidth}
         \centering
         \includegraphics[width=\linewidth]{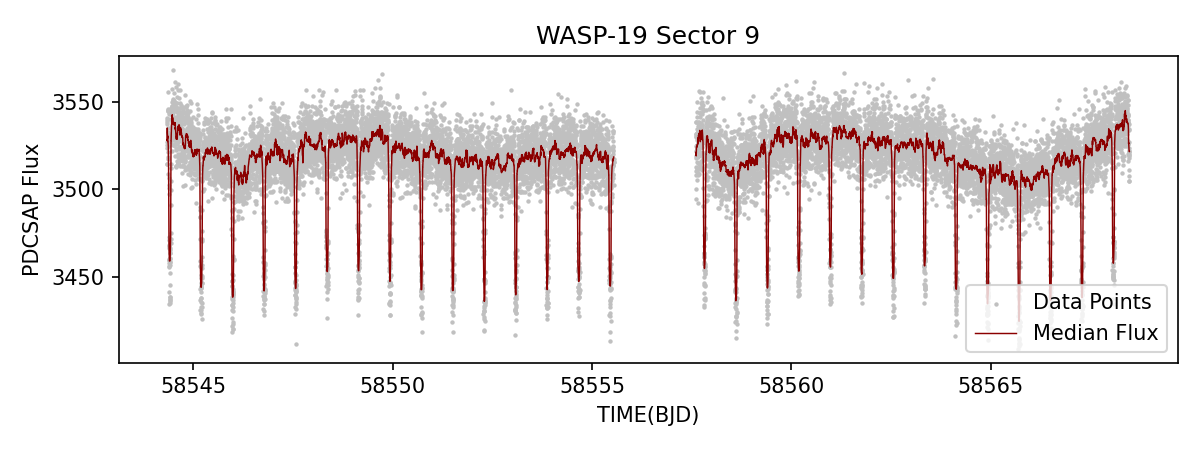}
         \label{wasp19_9}
     \end{subfigure}
     \begin{subfigure}[b]{\linewidth}
         \centering
         \includegraphics[width=\linewidth]{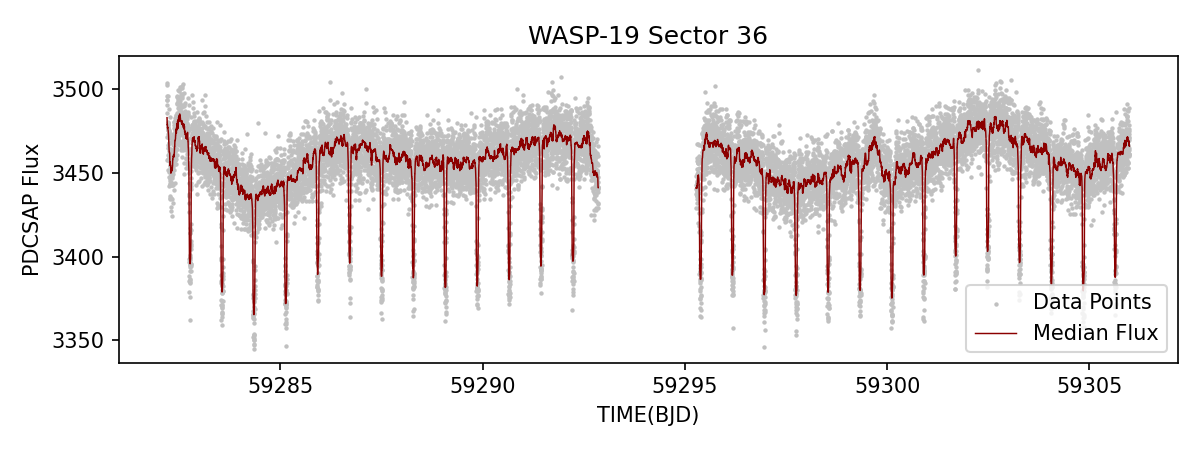}
         \label{wasp19_36}
     \end{subfigure}
        \caption{Plot of the PDC light curves of WASP-19 with NaNs and quality flagged points removed. Moving median filter over 32 points shown in red.  Long-term trends can be observed in both sectors, most likely due to stellar activity or uncorrected systematic errors.}
        \label{wasp19LC}
\end{figure}

\section{TESS data} 

\subsection{Observations}

WASP-18 (TIC 100100827) was observed by Camera 2 of TESS with a 2 minute cadence in Sector 2 (from 2018 August 22 to 2018 September 20), Sector 3 (from 2018 September 20 to 2018 October 18), Sector 29 (from 2020 August 26 to 2020 September 21), and Sector 30 (from 2020 September 23 to 2020 October 20). A total of 92 transits were extracted from the observations.\par

WASP-19 (TIC 35516889) was observed with the same cadence by Camera 2 of TESS in Sector 9 (from 2019 February 28 to 2019 March 26) and Sector 36 (from 2021 March 07 to 2021 April 01). We extracted 56 transits from the observations.

The TESS light curves were downloaded through the Mikulski Archive for Space Telescopes (MAST) portal\footnote{https://mast.stsci.edu/}. For the analysis performed in this work, we use the Presearch Data Conditioning Simple Aperture Photometry (PDCSAP) curves, which remove long-term non-astrophysical trends from the data and reduce systematic errors \citep{smith2012kepler, smith2017kepler}.\par

The data files include a quality parameter that indicates data points that have been affected by anomalous events such as attitude tweaks, momentum dumps, firing of thrusters, and other events \citep{tenenbaum2018tess}. We removed all non-zero quality flagged data points from our light curve to ensure reliable data and removed all NaN values caused by poor instrumental behaviour, but long-term trends are still present in the initial curves, especially in WASP-19 (Fig. \ref{wasp19LC}). This may be caused by the high stellar activity of WASP-19 \citep{hebb2009wasp, espinoza2019access, wong2020tess}, as well as some uncorrected systematic trends, and additional corrections are described in Sect. 3.2.\par

Some incomplete transits are seen in Figs. \ref{wasp18LC} and \ref{wasp19LC}, especially near the gaps in the middle which separate the two physical orbits from TESS. During the light curve processing described in Sect. 3.2, these incomplete transits were discarded, since the lack of a complete transit does not allow for a precise derivation of the transit timing \citep{barros2013transit}.

% =================================================================================

\subsection{Stellar activity correction}

As mentioned before, stellar activity is a likely cause of the trends shown in WASP-19 data (Fig. \ref{wasp19LC}). To take the effect of stellar activity and spots into account, we assume the maximum flux to correspond to the minimum amount of stellar spots. We applied a moving median filter over 32 points to each light curve and found the maximum median flux over all light curves in each target. We then used this value as the normalisation constant to obtain the relative flux. We also considered the maximum of each sector light curve for normalisation instead of the overall maximum, but the difference in the final result was not significant and we decided to keep the original approach. It is, however, important to note that this does not correct for spot occultation events \citep[e.g.][]{barros2013transit}. \par

The normalised curve was then split into individual transits. For that, we started by flagging the data points within three transit durations of the mid-transit time and by splitting the light curves into individual segments, removing the remaining data between transits, which allowed for an easier capture of short-term trends near the transits while minimising the influence of long-term trends. We applied a detrending model using a low-order polynomial and used the Bayesian information criterion (BIC) to find the optimal polynomial order for each segment. The vast majority of the individual light curves favoured a first-order polynomial trend, with only a few exceptions in WASP-18 favouring a second-order model. We normalised each individual transit light curve to an out of transit level of 1 using the obtained best-fit coefficients.

% =================================================================================
% SECTION 4 - TRANSIT ANALYSIS
% =================================================================================

\section{Transit analysis}

\begin{table}
\caption{System stellar parameters for each target, with WASP-18 parameters obtained from \cite{shporer2019tess} and WASP-19 parameters obtained from \cite{mancini2013physical}. Limb-darkening coefficients for the quadratic model ($u_1$ and $u_2$) were estimated with LDTk \citep{parviainen2015ldtk} using the $T_\mathrm{eff}$, $\log g$, and $\mathrm{[Fe/H]}$ from the references.}
\label{stellar}      % is used to refer this table in the text
\centering                          % used for centring table
\begin{tabular}{c c c}       
\hline\hline             
Parameter & WASP-18 & WASP-19\\    % table heading 
\hline
    $R_\star (R_\sun)$ & $1.26\pm0.04$ & $1.018\pm0.021$\\
    $M_\star (M_\sun)$ & $1.46\pm0.29$ & $0.935\pm0.045$\\
    $T_\mathrm{eff}$ (K) & $6431\pm48$ & $5460\pm90$\\      
    $\log g$ & $4.47\pm0.13$ & $4.432\pm0.013$\\
    $\mathrm{[Fe/H]}$ & $0.11\pm0.08$ & $0.14\pm0.11$\\
    $u_\mathrm{1}$ &  $0.361712\pm0.000917$ & $0.435405\pm0.001394$\\      
    $u_\mathrm{2}$ &  $0.158463\pm0.001641$ & $0.132505\pm0.002214$\\
\hline
\end{tabular}
\end{table}

The transits were modelled using the model from \texttt{batman} \citep{kreidberg2015batman} and we split our analysis into two parts. First we performed a fit of all transits aiming to obtain the transit parameters. We put together all individual segments already detrended and fit for reference time $T_\mathrm{0}$, period $P$, planet-to-star radius ratio $r_\mathrm{p}/R_\star$, scaled orbital semi-major axis $a/R_\star$, orbital inclination $i$, and the two limb-darkening coefficients of the quadratic model $u_\mathrm{1}$ and $u_\mathrm{2}$. Circular orbits with fixed $e=0$ and $\omega=90\degree$ were used in all analyses. The circularisation timescale for WASP-18b is around 15 Myr, while WASP-19b has a value around 2 Myr; both are at least an order of magnitude lower than the estimated system ages. Because of this, we expect the eccentricity to be zero in both planets, unless external perturbations occur. Nevertheless, we also fitted a model with varying eccentricity using Gaussian priors, and compared the fits using the BIC. We found that the circular models were strongly preferred (with a lower BIC)
over the eccentric models. Therefore, we assumed a circular model for all the remaining analyses in this paper.\par
We used a quadratic limb-darkening model with values estimated with the Limb-Darkening Toolkit \citep[LDTk,][]{parviainen2015ldtk} using the effective temperature ($T_\mathrm{eff}$), gravity ($\log g$), and metallicity ($\mathrm{[Fe/H]}$) of each star. The stellar parameters as well as the initial limb-darkening values can be found in Table \ref{stellar}.\par

\begin{figure}
     \centering
        \includegraphics[width=\linewidth]{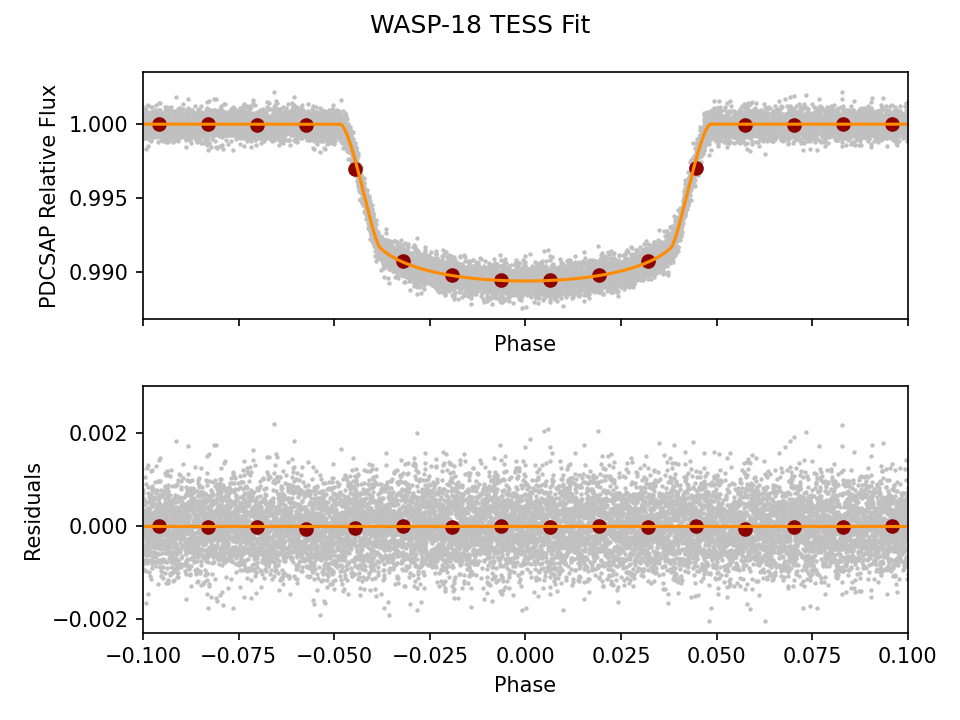}
        \caption{Best-fit model for WASP-18b phase-folded light curve of all TESS transits. }
        \label{w18best}
\end{figure}

\begin{figure}
     \centering
        \includegraphics[width=\linewidth]{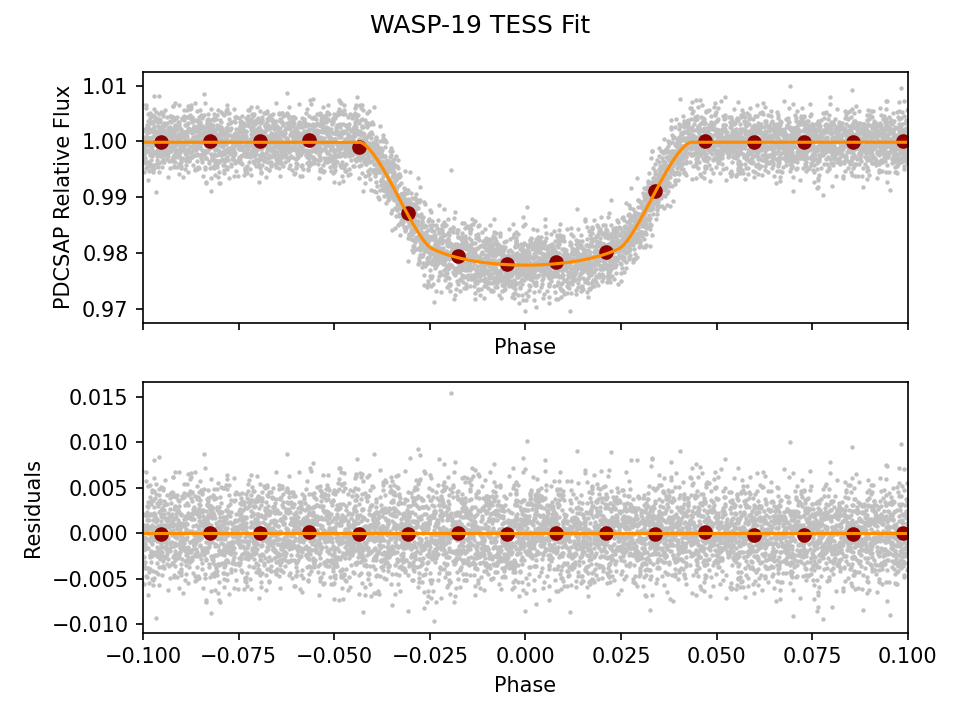}
        \caption{Best-fit model for WASP-19b phase-folded light curve of all TESS transits.}
        \label{w19best}
\end{figure}

\begin{table*}
\caption{System parameters and priors used in the transit modelling.}
\label{paramfit}      % is used to refer this table in the text
\centering                          % used for centring table
\begin{tabular}{c c c c} 
\hline\hline             
Parameter & Prior & Derived Value\\    % table heading 
\hline
    \textbf{WASP-18b}\\
    $T_\mathrm{0}$ (BJD) & $\mathcal{U}(2458021.125, 2458023.125)$ & $2458022.125237(37)$\\
    $P$ (days) & $\mathcal{U}(0.44145, 1.44145)$ & $0.941452397(44)$\\  
    $r_\mathrm{p}/R_\star$ & $\mathcal{U}(0.04776, 0.14776)$ & $0.09765\pm0.00017$\\  
    $a/R_\star$ & $\mathcal{U}(1.0, 5.0)$ & $3.456\pm0.017$\\  
    $i\ (\degree)$ & $\mathcal{U}(60, 90)$ & $83.58^{+0.25}_{-0.24}$\\
    $u_\mathrm{1}$ &  $\mathcal{N}(0.361712, 0.1)$ & $0.300\pm0.020$\\      
    $u_\mathrm{2}$ &  $\mathcal{N}(0.158463, 0.1)$ & $0.159\pm0.037$\\
    \hline
    \textbf{WASP-19b}\\
    $T_\mathrm{0}$ (BJD) & $\mathcal{U}(2456020.704, 2456022.704)$ & $2456021.70386(50)$\\  
    $P$ (days) & $\mathcal{U}(0.28884, 1.28884)$ & $0.78883896(14)$\\  
    $r_\mathrm{p}/R_\star$ & $\mathcal{U}(0.09233, 0.19233)$ & $0.14541^{+0.00070}_{-0.00077}$\\  
    $a/R_\star$ & $\mathcal{U}(1.0, 5.0)$ & $3.533^{+0.039}_{-0.037}$\\  
    $i\ (\degree)$ & $\mathcal{U}(60, 90)$ & $79.17^{+0.33}_{-0.31}$\\
    $u_\mathrm{1}$ &  $\mathcal{N}(0.435405, 0.1)$ & $0.403^{+0.061}_{-0.062}$\\      
    $u_\mathrm{2}$ &  $\mathcal{N}(0.132505, 0.1)$ & $0.118^{+0.084}_{-0.082}$\\
    \hline

\end{tabular}
\end{table*}

We used a Markov chain Monte Carlo (MCMC) algorithm based on the \texttt{emcee} ensemble sampler in Python \citep{foreman2013emcee} for the transit analysis. We defined a number of walkers of 4 times the number of fitted parameters over 10000 steps and discarded the burn-in before retrieving the results from the posterior distributions. We defined uniform priors for all parameters except the limb-darkening coefficients, for which we defined Gaussian priors centred in the values from LDTk with a conservative standard deviation of 0.1. 
The results of the transit analysis and the best-fit parameters are listed in Table \ref{paramfit} together with the used priors. We obtained updated planetary and transit parameters for both targets with good agreement with the parameters obtained previously.\par

For the second part of the analysis, we fitted the mid-transit times of each transit separately. We fixed the previously obtained value for $P$, $u_\mathrm{1}$, and $u_\mathrm{2}$, allowing $r_\mathrm{p}/R_\star$, $a/R_\star$, and $i$ to float with Gaussian priors defined by the posteriors from the previous analysis. A uniform prior was used for the mid-transit time $T_\mathrm{mid}$. We used the same MCMC approach as for the first part of the analysis, with a number of walkers equal to 4 times the number of free parameters over 5000 steps for each transit, discarding burn-in, and included a first-degree polynomial trend in the fit to propagate detrending errors in all curves. In this analysis we obtained precise timings for all TESS transits currently available, which are listed in Tables \ref{w18_times} and \ref{w19_times} together with previously obtained transits and their origin paper. \par

Sectors 2 and 3 of WASP-18 have been analysed previously by \cite{shporer2019tess} and \cite{maciejewski2020planet}. Our results agree with the ones from \cite{maciejewski2020planet} within 0.5-1$\sigma$ for the period and radius and 2$\sigma$ for $a/R_\star$ and inclination. \cite{shporer2019tess} considered a non-zero eccentricity for their analysis and did not fit the limb-darkening parameters, which leads to a more significant deviation from our results, with a 3$\sigma$ agreement in radius but a deviation of 2-3\% in $a/R_\star$ and inclination.\par

The radius of WASP-19b significantly deviates from previous works. \cite{espinoza2019access} compare their results with the ones from \cite{hebb2009wasp}, \cite{tregloan2013transits}, and \cite{mancini2013physical} with good agreement; however, \cite{wong2020tess} used sector 9 of TESS and found an increase of 10-15\% in the transit depth, reporting a value of 23240 ppm against 20258 ppm as reported by \cite{espinoza2019access}. We ran individual fits of sectors 9 and 36 and found that between both sectors there is a significant difference of around 3\% in depth, with 21590 ppm in sector 9 and 20996 ppm in sector 36. While our transit depth estimation is not as far from the previous ones as the one from \cite{wong2020tess}, we still find an increase of over 4\% in depth comparing our value with \cite{espinoza2019access} using both TESS sectors (21145 ppm), and over 6\% if we use sector 9 only. \cite{wong2020tess} suggest that an increase in stellar spots on the surface during sector 9 measurements is the cause for the apparently bigger radius. Our smaller difference when only using sector 9 data may therefore arise from our initial correction with individual transit segments and posterior normalisation, which was aimed to minimise the effect of spots. The different values obtained for sector 36 and the fact that the remaining parameters all agree with \cite{espinoza2019access} in <1$\sigma$ levels suggests stellar activity is indeed high and is causing significant variations in the transit depths when using TESS data.

% =================================================================================
% SECTION 5 - ORBITAL DECAY ANALYSIS
% =================================================================================

\section{Timing and orbital decay analysis}

The main goal of our paper is to better constrain tidal decay in promising targets. We used our derived transit times from TESS together with other transit times from the literature to look for TTVs in WASP-18b and WASP-19b. All timings are listed in Tables \ref{w18_times} and \ref{w19_times}, referencing the original paper where the transit was first reported. However, \cite{maciejewski2020planet}, while compiling all previous transits of WASP-18b, ran new photometric analyses of the majority of the reported transits. Several transits from TRESCA used by \cite{petrucci2020discarding} were excluded from the WASP-19 TTV analysis. While they correspond to recent observations, we verified the normalised residuals of those times against a constant period model and found a high amount of scatter and high uncertainties when compared to TESS transits. Since the data are contemporary with the more precise TESS transits, we decided not to include them in the analysis.\par

In order to study the period variations in both targets, we estimated the time of each individual transit as a function of the epoch using a linear and a quadratic model. The linear model assumes a circular orbit with a constant period as
\begin{equation}
        T_\mathrm{mid}=T_\mathrm{0}+P\times E
\end{equation}
where $T_\mathrm{0}$ is the reference mid-transit time ($E=0$), $P$ is the orbital period, and $E$ is the epoch number counted from $T_\mathrm{0}$. The quadratic model includes an additional term, taking a uniform period change into account as
\begin{equation}
        T_\mathrm{mid}=T_\mathrm{0}+P\times E+\frac{1}{2}\dot{P}\times P\times E^2
\end{equation}
where $\dot{P}=dP/dT$ is the rate of change in orbital period given by its derivative over time.\par
Comparing the two models using the BIC allowed us to look for a significant detection of a period change. The derivative over time is related to the star tidal quality factor $Q_\star$, which corresponds to the rate of energy dissipation due to tidal interactions \citep{goldreich1966q}. The larger the value of $Q_\star$, the lower the tidal dissipation efficiency is. We used a parametrisation for $Q_\star$ from the tidal interaction model by \cite{zahn1977tidal}
\begin{equation}
        Q'_\star=\frac{3Q_\star}{2k_\star}
\end{equation}
where $k_\star$ is the stellar Love number, which measures how much of the star's density is concentrated in its core and is related to its response to tidal forces. This leads to a relationship between $Q'_\star$ and $\dot{P}$
\begin{equation}
        Q'_\star=-\frac{27}{2}\pi\left(\frac{M_\mathrm{p}}{M_\star}\right)\left(\frac{a}{R_\star}\right)^{-5}\dot{P}^{-1}
        \label{qstar}
\end{equation}
where $M_\mathrm{p}/M_\star$ is the planet-star mass ratio and $a/R_\star$ is the scaled orbital semi-major axis. Planets similar to WASP-18b, with a high planet-star mass ratio, are very good candidates for tidal decay studies.\par 
We assumed Gaussian priors for $T_\mathrm{0}$ and $P$, using the previous best-fit values and uncertainties, and a uniform prior for $\dot{P}$, $\mathcal{U}(-10, 2)$. We ran an MCMC analysis with the \texttt{emcee} code using all available transit times $T_\mathrm{mid}$ to fit for the value of $\dot{P}$ and refine $T_\mathrm{0}$ and $P$. We ran the model over 32 walkers with 10000 steps each, discarding burn-in.

% =================================================================================
% SECTION 6 - RESULTS AND DISCUSSION
% =================================================================================

\section{Results and discussion}

\begin{figure*}
     \centering
        \includegraphics[width=\linewidth]{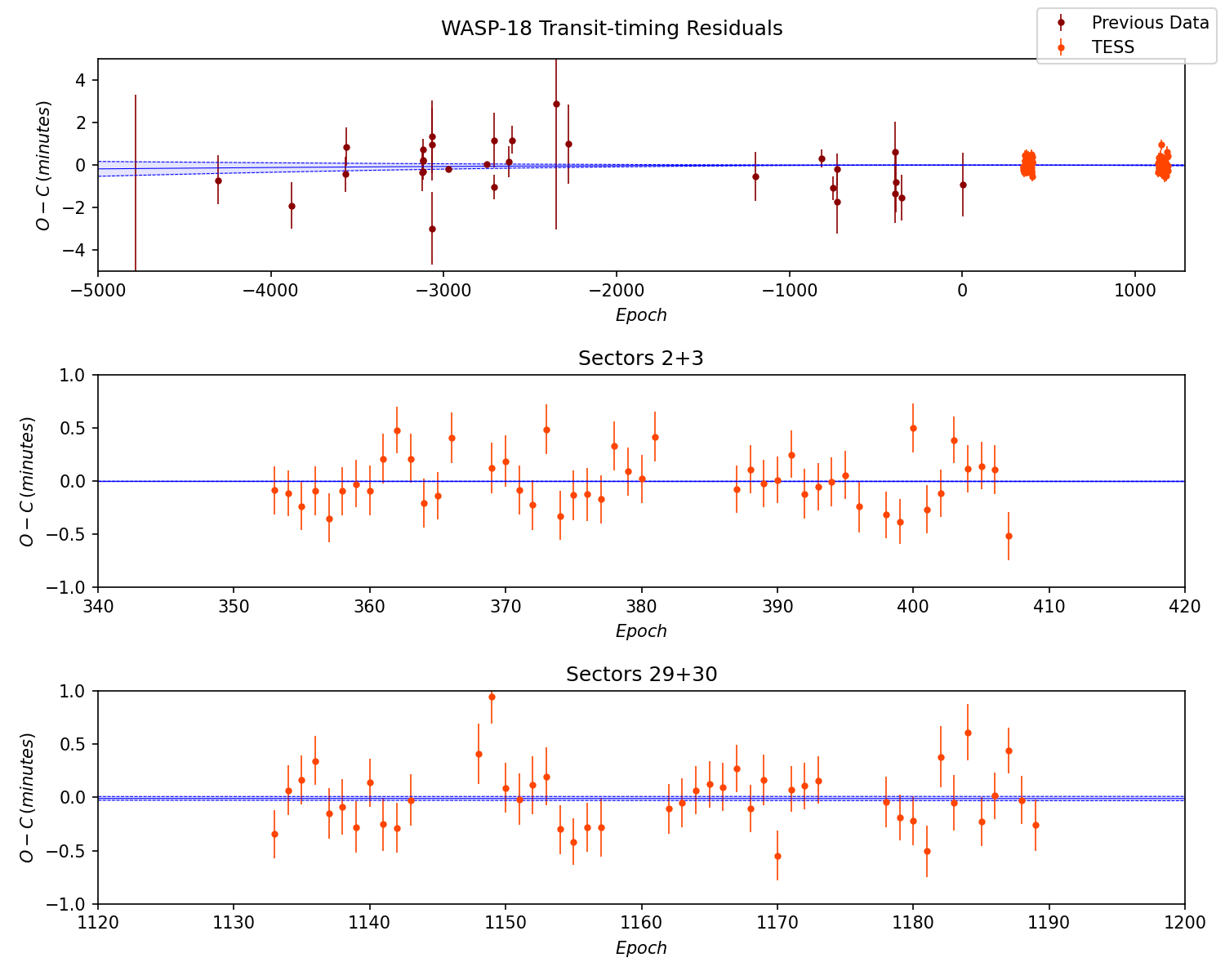}
        \caption{Transit difference (O-C) against the epoch, counted from the initial reference time for WASP-18b. Previous results from the literature (see Table \ref{w18_times} for reference) are shown in dark red while our results from TESS are shown in orange. The blue line shows the fitted TTV model with the blue shaded area showing the error margin for $\dot{P}$. The residuals show no signs of additional small-scale structures and most are compatible with zero within 1$\sigma$. Upper plot: Complete view of all the transits used in the fit. Middle plot: Zoom of the transit timings of TESS Sectors 2 and 3. Bottom plot: Zoom of the transit timings of TESS Sectors 29 and 30.}
        \label{w18pdot}
\end{figure*}

\begin{figure*}
     \centering
        \includegraphics[width=\linewidth]{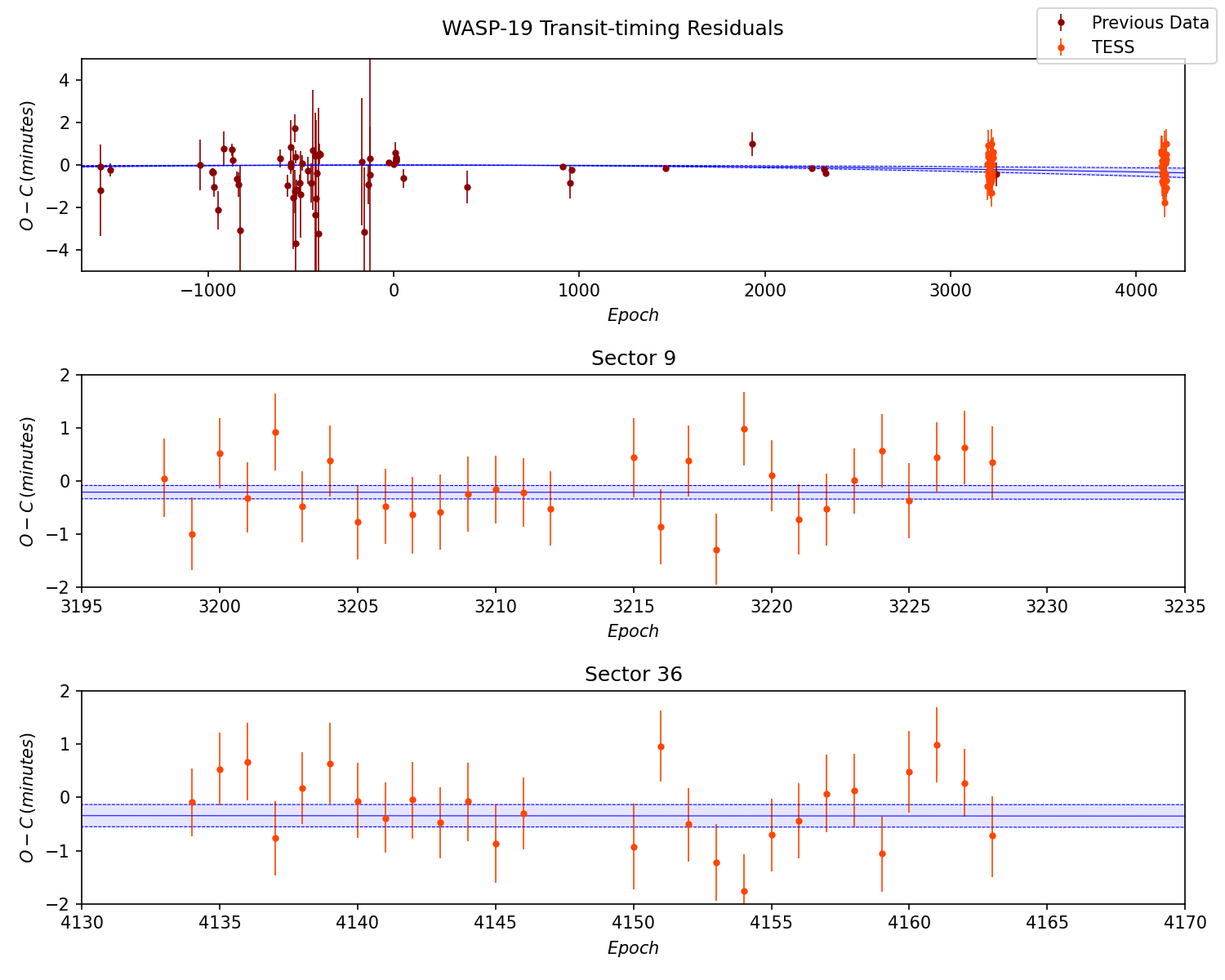}
        \caption{Same as Figure \ref{w18pdot}, but for WASP-19b. The points from previous data are plotted in dark red and listed with references in Table \ref{w19_times}. As with WASP-18b, we did not find any additional signals in the residuals, with most being compatible with zero at 1$\sigma$. Upper plot: Complete view of all the transits used in the fit. Middle plot: Zoom of the transit timings of TESS Sector 9. Bottom plot: Zoom of the transit timings of TESS Sector 36.}
        \label{w19pdot}
\end{figure*}

The results for the TTV fits are listed in Table \ref{ttvresults}. We compared the BIC values of the linear and quadratic models, with the linear model being favoured for both targets (-4.09 against 0.73 for WASP-18b and -3.91 against 0.85 for WASP-19b), suggesting there is still no significant evidence towards a change in orbital period. The best-fit quadratic models are plotted in Figs. \ref{w18pdot} and \ref{w19pdot}. 

\begin{figure}
     \centering
        \includegraphics[width=\linewidth]{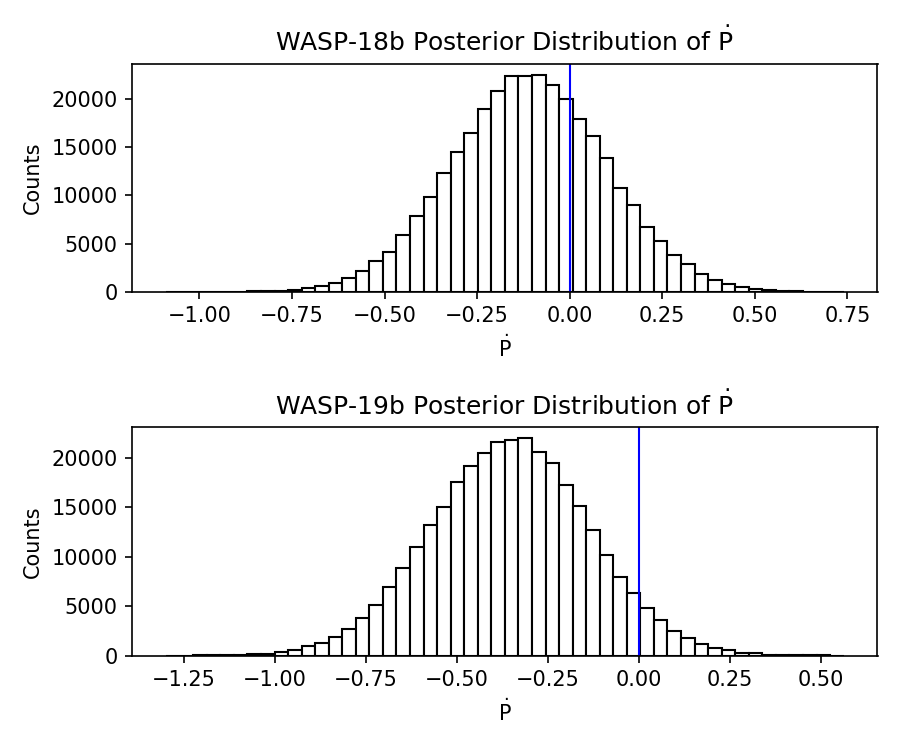}
        \caption{Plot of the posterior distribution of $\dot{P}$ from the quadratic ephemerides fit of WASP-18b (upper) and WASP-19b (lower) TTVs. While it is clear the distribution of WASP-18 is centred at a non-zero value, it still includes zero in its 68\% confidence interval. On the other hand, in the distribution for WASP-19b, we see zero is absent from the 1$\sigma$ confidence level, but present at a 2$\sigma$ level.}
        \label{w18pdf}
\end{figure}

\begin{table*}
\caption{Best-fit results for our TTV analysis, with the linear constant period model and the quadratic model of orbital decay. We note that $Q'_\star$ is given as a lower limit with a 95\% confidence level with the corresponding upper limit of $\dot{P}$.}
\label{ttvresults}      % is used to refer this table in the text
\centering
% used for centring table
\begin{tabular}{c c c}       
\hline\hline
 & \multicolumn{2}{l}{Constant Period}\\  
Target & $T_\mathrm{0}$ (BJD) & $P$ (days)\\
\hline
WASP-18 & 2458022.125226(16) & 0.941452407(14)\\
WASP-19 & 2456021.703698(20) & 0.788839004(13)\\
\hline
\vspace{1mm}
\end{tabular}
\begin{tabular}{c c c c c c}      
\hline\hline
 & \multicolumn{5}{l}{Decay Quadratic Model}\\  
Target & $T_\mathrm{0}$ (BJD) & $P$ (days) & $\dot{P}$ & $\dot{P}$ upper limit & $Q'_\star$ lower limit\\
\hline
WASP-18 & 2458022.125234(22) & 0.941452400(20) & $(-0.11\pm0.21)\times10^{-10}$ & $<-0.45\times10^{-10}$ & $>(1.42\pm0.34)\times10^7$\\
WASP-19 & 2456021.703704(21) & 0.788839038(25) & $(-0.35\pm0.22)\times10^{-10}$ & $<-0.71\times10^{-10}$ & $>(1.26\pm0.10)\times10^6$\\
\end{tabular}
\end{table*}

\begin{table*}[h]
\caption{Comparison of $\dot{P}$ and $Q'_\star$ values from recent works. All upper or lower limits were evaluated at a 95\% confidence level (2$\sigma$).}
\label{qstarcomp}      % is used to refer this table in the text
\centering                          % used for centring table
\begin{tabular}{c c c}       
\hline\hline             
Source & $\dot{P}$ & $Q'_\star$ \\    % table heading 
\hline 
    \textbf{WASP-18b} \\
    This work & $(-0.11\pm0.21)\times10^{-10}$ & $>(1.42\pm0.34)\times10^7$\\
    \cite{maciejewski2020planet} & $(0.11\pm1.17)\times10^{-10}$ & $>3.9\times10^6$ \\
    \cite{patra2020continuing} &  $(-1.2\pm1.3)\times10^{-10}$ & $>(1.7\pm0.4)\times10^6$ \\
    \cite{shporer2019tess} & $(-0.60\pm2.18)\times10^{-10}$ & $>1.73\times10^6$\\
\hline
    \textbf{WASP-19b} \\
    This work & $(-0.35\pm0.22)\times10^{-10}$ & $>(1.26\pm0.10)\times10^6$\\
    \cite{petrucci2020discarding} & $<-0.73\times10^{-10}$ & $>(1.23\pm0.231)\times10^6$ \\
    \cite{patra2020continuing} &  $(-2.06\pm0.42)\times10^{-10}$ & $=(5.0\pm1.5)\times10^5$ \\
\hline
\end{tabular}
\end{table*}

We see from Table \ref{ttvresults} and Fig. \ref{w18pdf} that the $\dot{P}$ value for WASP-18b is still far from showing evidence of decay, being comfortably compatible with zero at 1$\sigma$ and thus favouring a constant period. The most recent studies of WASP-18b by \cite{shporer2019tess}, \cite{maciejewski2020planet}, and \cite{patra2020continuing} report values consistent with these findings, with the ones of the latter being the closest to a varying period with $\dot{P}= (-1.2\pm1.3)\times10^{-10}$. WASP-18b has a relatively high planet-star mass ratio, which makes it one of the main targets for observing tidal decay. \cite{wilkins2017searching} estimated a minimum of $Q'_\star > 10^{6}$ at 95\% confidence; this value was only slightly improved by \cite{shporer2019tess} and \cite{maciejewski2020planet} using TESS data (see Table \ref{qstarcomp}). We are able to improve upon the minimum value of $Q'_\star$ with the additional transits from Sectors 29 and 30 by an order of magnitude compared to previous works. Despite the increased precision, we did not detect evidence of tidal decay in the orbit of WASP-18b and so we derived a lower limit of $Q'_\star>(1.42\pm0.34)\times10^7$ at 95\% confidence, corresponding to a $\dot{P}$ upper limit of $-0.45\times10^{-10}$. From our result of $Q'_\star$, the estimated spiral in-fall timescale is around $5\times 10^{7}\ \mathrm{yr}$, which is longer than originally expected. For comparison, WASP-12b, the only planet with confirmed tidal decay so far has a corresponding $Q'_\star = (1.50\pm0.11)\times10^5$ \citep{wong2022tess}, which is two orders of magnitude lower than the minimum value we found for WASP-18. WASP-12 is also an F-type star, with an effective temperature of 6150 K, with a mass similar to WASP-18. If we assume Eq. (4) to be correct and $Q'_\star$ to be the same for the same type of stars, WASP-18b should have a faster decay than WASP-12b, due to the planet-star mass ratio (which is 10 times higher). The absence of significant decay in WASP-18b so far shows that despite the similar stellar type and mass, $Q'_\star$ is significantly different between the systems, suggesting the tidal dissipation processes may differ within stars of a similar type. \par

% ===========================================================
% WASP-19
% ===========================================================

WASP-19b shows a decay rate different from zero at a 1$\sigma$ level (Table \ref{ttvresults} and Fig. \ref{w18pdf}). Comparing our results with the ones obtained by \cite{patra2020continuing}, our value of $Q'_\star$ is higher than theirs by almost an order of magnitude, with our $\dot{P}$ absolute value being almost 6 times smaller, deviating from theirs by more than 7$\sigma$. Since we are unable to confirm the detection of a period change, we derived a lower limit for $Q'_\star$ of WASP-19 of $(1.26\pm0.10)\times10^6$ at a 95\% confidence level, corresponding to an upper limit of $\dot{P} < -0.71\times10^{-10}$.
\cite{petrucci2020discarding} include additional transits that appear to discard a change in the period of WASP-19b. As mentioned before, we opted to leave this data set out of our analysis due to the high scatter in the transit times and the higher normalised residuals we found in them, with several outliers showing up near the TESS observations. We opted to include only the new data from TESS for this reason. It is important to note that \cite{petrucci2020discarding} derived a value of $\dot{P}$ that favours the absence of orbital decay, listing an upper limit for the decay rate and a lower limit for $Q'_\star$ (see Table \ref{qstarcomp}).
These inconsistent results and the high stellar activity of WASP-19 make it a challenge to obtain significant conclusions. As we have discussed before, the transit depth found in TESS transits is significantly higher than reported before by previous authors \citep{hebb2009wasp, tregloan2013transits, mancini2013physical, espinoza2019access}. \cite{wong2020tess} suggest that star spots are at fault for this difference, but they were unable to find spot-crossing events in individual transits. The effect has also been studied in detail by \cite{tregloan2013transits}, who found anomalies in two consecutive light curves due to spots. \cite{espinoza2019access} claim that WASP-19 shows rotational variability that is caused by giant spots that impact the observations. \cite{patra2020continuing} suggest that there are indications that the stellar activity is causing systematic errors affecting transit times, which is further addressed by \cite{petrucci2020discarding}, citing the uncertainty in the inclination parameter as the main cause of errors in transit times. Therefore, while the results may seem promising, the uncertainties caused by stellar activity indicate that one should use caution regarding the confirmation or when ruling out tidal decay in WASP-19. The system is expected to be observed by the CHaracterising ExOplanet Satellite (CHEOPS) soon, which may shed some new light on the physical processes behind this system.

% =================================================================================
% OTHER TARGETS
% =================================================================================

\begin{table*}
\caption{Summary of the status of the main targets of orbital decay studies from the literature. Minimum values of $Q'_\star$ given at a 95\% confidence level, with the exception of WASP-103 which has a confidence level of 99.7\%.}
\label{status}      % is used to refer this table in the text
\centering
% used for centring table
\resizebox{\textwidth}{!}{
\begin{tabular}{c c c c c c c c}       
\hline\hline
Target & Type & $T_\mathrm{eff}$ (K) & Age (Gyr) & $M_\star/M_\sun$(a) & $\dot{P}$ & $Q'_\star$ & Reference\\
\hline
WASP-18 & F & 6431 & $\approx1$ & 1.46(29) & $(-0.10\pm0.21)\times10^{-10}$ & $>(1.42\pm0.34)\times10^7$ & This work\\
WASP-19 & G & 5460 & $\approx10$ & 0.935(41) & $(-0.35\pm0.22)\times10^{-10}$ & $>(1.26\pm0.10)\times10^6$ & This work\\
WASP-12 & F & 6150 & $\approx2$ & 1.434(11) & $(-9.45\pm0.30)\times10^{-10}$ & $=(1.50\pm0.11)\times10^5$ & \cite{wong2022tess}\\
WASP-4 & G & 5400 & $\approx7$ & $1.37(18)$ (b) & $(-7.33\pm0.23)\times10^{-10}$ & $=(5.1\pm0.9)\times10^4$ & \cite{turner2021characterizing}\\
WASP-43 & K & 4195 & $>0.2$ & 0.717(25) & $(-1.11\pm0.22)\times10^{-10}$ & $>(4.01\pm1.15)\times10^5$ & \cite{davoudi2021investigation}\\
WASP-103 & F & 6013 & $\approx5$ & 1.21(11) & $(3.6\pm1.6)\times10^{-10}$ & $>(3.3\pm1.15)\times10^6$(c) & \cite{barros2022detection}\\
KELT-16 & F & 6250 & $\approx3$ & 1.211(46) & $(-3.36\pm4.15)\times10^{-10}$ & $>(1.9\pm0.8)\times10^5$ & \cite{mancini2021}\\
\hline
\multicolumn{8}{l}{(a) \cite{patra2020continuing}}\\
\multicolumn{8}{l}{(b) \cite{bouma2019wasp}}\\
\multicolumn{8}{l}{(c) 99.7\% confidence level}\\

\end{tabular}}

\end{table*}

\section{Tidal decay in other systems}

There are several other targets that are interesting for tidal decay studies due to the characteristics of their stars and the predicted tidal decay rate they present. As mentioned before, WASP-12b has already been confirmed as having a decaying period due to tides and WASP-4b has also shown signs of a rapid change in period; although, the reason for that change is still unclear \citep{turner2021characterizing}. Systems such as WASP-43 \citep{hellier2011wasp}, WASP-103 \citep{gillon2014wasp}, and KELT-16 \citep{oberst2017kelt} have also been subject of tidal decay analyses. In Table \ref{status}, we present a summary of the current status of orbital decay studies for these targets and a comparison with our results.\par

WASP-18 presents a lower limit of $Q'_\star$ that is higher than the measured value for WASP-12 and WASP-4, indicating that the orbit of WASP-18b might be decaying slower. Some systems have a minimum value of $\approx10^5$, which is consistent with the measured value in WASP-12. According to the values we collected, KELT-16 and WASP-43 appear to be good candidates to show tidal decay. WASP-43 has the added interest of being a K-type star, unlike the other promising targets, making it an important test subject for the measurement of $Q'_\star$ in colder stars. We note that WASP-103b has been recently found to show a hint of an increase in period, with new transits obtained by CHEOPS indicating that perhaps tidal decay is not the dominant factor in the period change of the planet \citep{barros2022detection}. It is still possible to obtain a lower limit for $Q'_\star$, shown in Table \ref{status}, since a negative $\dot{P}$ cannot be excluded within a 99.7\% confidence level. The minimum $Q'_\star$ found by \citet{barros2022detection} is higher than previously found \citep[e.g.][]{patra2020continuing}.
WASP-4b shows the fastest decay to date; however, the $Q'_\star$ value it shows is lower than expected by theory, making it still uncertain if the change in period is caused by tidal decay or not. \par

Theoretically, one could assume that the same type of stars would share similar values of $Q'_\star$. However, the different values found, differing by as much as two orders of magnitude, indicate that even within stars of similar spectral types and ages similar to WASP-12 and WASP-18, stellar physics may vary significantly. The better constraining of $Q'_\star$ is crucial to address the reasons behind the different $Q'_\star$ values and to give us new information about stellar interiors.

\section{Summary}

We performed an analysis of two hot Jupiters expected to show signs of orbital decay due to a tidal interaction: WASP-18b and WASP-19b. We analysed all the available sectors from TESS obtained for both targets between 2019 and 2021 and obtained updated planet parameters. We refined the planetary and transit parameters obtaining a good agreement with previous reports for WASP-18b. For WASP-19b, our parameters agree with \cite{espinoza2019access} at a 1$\sigma$ level; however, we found that the transit depth (and the planet radius) differs significantly with TESS data, with an increase of 6\%. This is likely to be caused by stellar variability as proposed by several authors. We attempted to correct the light curves to account for the effect of the non-occulted star spots but, nevertheless, the transit depth varies even between the two TESS sectors we analysed (3\%).\par
We further obtained transit times for all the available TESS observations and used our results, together with previous timings, to look for signs of orbital decay. We found no statistically significant evidence of decay in either targets, with the linear model being favoured by the BIC. On WASP-18b, we derived a new minimum value of $Q'_\star$ of $(1.42\pm0.34)\times10^7$ at 2$\sigma$, which is two orders of magnitude higher that the value of WASP-12. We found a negative $\dot{P}$ within $1\sigma$ for WASP-19b, but the results are likely influenced by possible systematic errors induced by stellar activity that is unaccounted for and they are not statistically significant. We derived a lower limit for $Q'_\star$ of $(1.26\pm0.10)\times10^6$ at 2$\sigma$.\par
We found that $Q'_\star$ ranges from at least between $10^5$ and $10^7$ for selected F- and G-type stars, indicating that tidal dissipation works differently even within similar stars. Most of the mentioned targets are to be observed by TESS and CHEOPS in the future and those observations may be able to help improve the constrains on $Q'_\star$ and find out more about the tidal decay process and the physics involved in tidal dissipation.

% =================================================================================

\begin{acknowledgements}
This work was supported by FCT – Fundação para a Ciência e a Tecnologia through national funds and by FEDER through COMPETE2020 – Programa Operacional Competitividade e Internacionalizacão by these grants: UID/FIS/04434/2019, UIDB/04434/2020, UIDP/04434/2020, PTDC/FIS-AST/ 32113/2017 and POCI-01-0145-FEDER-032113; PTDC/FIS-AST/28953/2017 and POCI-01-0145-FEDER-028953; PTDC/FIS-AST/28987/2017 and POCI-01-0145-FEDER-028987; UIDB/04564/2020, UIDP/04564/2020, PTDC/FIS-AST/7002/2020, POCI-01-0145-FEDER-022217 and POCI-01-0145-FEDER-029932. N.M.R. acknowledges support from FCT through grant DFA/BD/5472/2020. O.D.S.D. is supported in the form of work contract (DL 57/2016/CP1364/CT0004) funded by national funds through FCT.
This paper includes data collected by the TESS mission. Funding for the TESS mission is provided by the NASA’s Science Mission Directorate. We acknowledge the use of public TESS data from pipelines at the TESS Science Office and at the TESS Science Processing Operations Center. Resources supporting this work were provided by the NASA High-End Computing (HEC) Program through the NASA Advanced Supercomputing (NAS) Division at Ames Research Center for the production of the SPOC data products.
Some of the data presented in this paper were obtained from the Mikulski Archive for Space Telescopes (MAST). STScI is operated by the Association of Universities for Research in Astronomy, Inc., under NASA contract NAS5-26555. Support for MAST for non-HST data is provided by the NASA Office of Space Science via grant NNX13AC07G and by other grants and contracts.
\end{acknowledgements}

% =================================================================================

\bibliographystyle{aa} 
\bibliography{refs} 

% =================================================================================

\clearpage
\onecolumn

\begin{appendix}
\section{Additional tables}

\begin{longtable}{r l l l l l}
\caption{WASP-18b transit times with epochs calculated from reference time $T_\mathrm{0}=2458022.1249563$. Values marked with an asterisk were derived by \cite{maciejewski2020planet} using the source light curve.}
\label{w18_times} \\     % is used to refer this table in the text

\hline\hline             
Epoch & $T_\mathrm{mid}$ (BJD) & \multicolumn{2}{c}{Residuals} & Band-pass & Light Curve Reference \\    % table heading 
\hline 
\endfirsthead  

\multicolumn{6}{c}
{{\tablename\ \thetable{} -- continued}} \\
\hline Epoch & $T_\mathrm{mid}$ (BJD) & \multicolumn{2}{c}{Residuals} & Band-pass & Light Curve Reference \\  \hline 
\endhead

\hline \multicolumn{6}{r}{{Continues on the next page}} \\ \hline
\multicolumn{6}{l}{{$^*$ Value derived by \cite{maciejewski2020planet}}}
\endfoot

\hline
\endlastfoot

$-10150$ & $2448466.365700^*$ & $+0.0328^*$ & $-0.0209^*$ & \textit{I} & Hipparcos, ESA (1997)\\
$-4784$ & $2453518.21258^*$ & $+0.00842^*$ & $-0.00666^*$ & \textit{I} & \cite{pojmanski1997all}\\
$-4304$ & $2453970.113611^*$ & $+0.000782^*$ & $-0.000815^*$ & $0.4-0.7\ \mu$m & \cite{butters2010first}\\
$-3881$ & $2454368.347146^*$ & $+0.000748^*$ & $-0.00077^*$ & $0.4-0.7\ \mu$m & \cite{butters2010first}\\
$-3566$ & $2454664.905676^*$ & $+0.000582^*$ & $-0.000576^*$ & $0.4-0.7\ \mu$m & \cite{hellier2009orbital}\\
$-3565$ & $2454665.84803^*$ & $+0.000629^*$ & $-0.00064^*$ & $0.4-0.7\ \mu$m & \cite{hellier2009orbital}\\
$-3122$ & $2455082.910597^*$ & $+0.000598^*$ & $-0.000634^*$ & \textit{V} & \cite{southworth2009physical}\\
$-3121$ & $2455083.852439^*$ & $+0.000315^*$ & $-0.000302^*$ & \textit{V} & \cite{southworth2009physical}\\
$-3120$ & $2455084.793919^*$ & $+0.000243^*$ & $-0.000244^*$ & \textit{V} & \cite{southworth2009physical}\\
$-3119$ & $2455085.734997^*$ & $+0.000267^*$ & $-0.000263^*$ & \textit{V} & \cite{southworth2009physical}\\
$-3118$ & $2455086.677153^*$ & $+0.000344^*$ & $-0.00035^*$ & \textit{V} & \cite{southworth2009physical}\\
$-3068$ & $2455133.7472^*$ & $+0.001181$ & $-0.001181$ & \textit{I} & \cite{cortes2020tramos}\\
$-3067$ & $2455134.6914^*$ & $+0.001181$ & $-0.001181$ & \textit{I} & \cite{cortes2020tramos}\\
$-3066$ & $2455135.6331^*$ & $+0.001181$ & $-0.001181$ & \textit{I} & \cite{cortes2020tramos}\\
$-2975$ & $2455221.3042$ & $+0.0001$ & $-0.0001$ & $3.6\ \mu$m & \cite{maxted2013spitzer}\\
$-2751$ & $2455432.1897$ & $+0.0001$ & $-0.0001$ & $4.5\ \mu$m & \cite{maxted2013spitzer}\\
$-2710$ & $2455470.7885$ & $+0.0004$ & $-0.0004$ & \textit{I+z} & \cite{maxted2013spitzer}\\
$-2707$ & $2455473.6144$ & $+0.0009$ & $-0.0009$ & \textit{I+z} & \cite{maxted2013spitzer}\\
$-2621$ & $2455554.5786$ & $+0.0005$ & $-0.0005$ & \textit{I+z} & \cite{maxted2013spitzer}\\
$-2604$ & $2455570.584$ & $+0.00045$ & $-0.00048$ & \textit{I+z} & \cite{maxted2013spitzer}\\
$-2348$ & $2455811.597^*$ & $+0.004097$ & $-0.004097$ & \textit{I} & \cite{cortes2020tramos}\\
$-2279$ & $2455876.5559$ & $+0.0013$ & $-0.0013$ & \textit{z'} & \cite{maxted2013spitzer}\\
$-1196$ & $2456896.1478$ & $+0.0008$ & $-0.0008$ & $1.1-1.7\ \mu$m & \cite{wilkins2017searching}\\
$-814$ & $2457255.7832$ & $+0.0003$ & $-0.00029$ & \textit{z} & \cite{wilkins2017searching}\\
$-746$ & $2457319.801$ & $+0.00039$ & $-0.00038$ & \textit{z} &\cite{wilkins2017searching}\\
$-726$ & $2457338.6296^*$ & $+0.00106^*$ & $-0.00108^*$ & \textit{r'} &\cite{kedziora2019secondary}\\
$-725$ & $2457339.572103^*$ & $+0.000516^*$ & $-0.000506^*$ & \textit{r'} & \cite{kedziora2019secondary}\\
$-388$ & $2457656.84078$ & $+0.000972$ & $-0.000972$ & \textit{I} & \cite{cortes2020tramos}\\
$-387$ & $2457657.78359$ & $+0.000972$ & $-0.000972$ & \textit{I} & \cite{cortes2020tramos}\\
$-386$ & $2457658.72404$ & $+0.000972$ & $-0.000972$ & \textit{I} & \cite{cortes2020tramos}\\
$-353$ & $2457689.79147^*$ & $+0.00075$ & $-0.00075$ & \textit{z'} & \cite{patra2020continuing}\\
$5$ & $2458026.83186^*$ & $+0.001042$ & $-0.001042$ & \textit{R} & \cite{cortes2020tramos}\\
$353$ & $2458354.45787421$ & $+0.0001575586$ & $-0.0001564676$ & $0.6-1.0\ \mu$m & TESS S2\\
$354$ & $2458355.39930413$ & $+0.0001488504$ & $-0.0001511559$ & $0.6-1.0\ \mu$m & TESS S2\\
$355$ & $2458356.34067383$ & $+0.0001539404$ & $-0.0001580069$ & $0.6-1.0\ \mu$m & TESS S2\\
$356$ & $2458357.28222453$ & $+0.0001567581$ & $-0.0001626202$ & $0.6-1.0\ \mu$m & TESS S2\\
$357$ & $2458358.22349873$ & $+0.0001580067$ & $-0.0001633241$ & $0.6-1.0\ \mu$m & TESS S2\\
$358$ & $2458359.16513058$ & $+0.0001620091$ & $-0.0001548205$ & $0.6-1.0\ \mu$m & TESS S2\\
$359$ & $2458360.10662774$ & $+0.0001524249$ & $-0.0001600427$ & $0.6-1.0\ \mu$m & TESS S2\\
$360$ & $2458361.04803607$ & $+0.0001606493$ & $-0.0001632962$ & $0.6-1.0\ \mu$m & TESS S2\\
$361$ & $2458361.98969784$ & $+0.000163606$ & $-0.000161472$ & $0.6-1.0\ \mu$m & TESS S2\\
$362$ & $2458362.93133366$ & $+0.0001485966$ & $-0.0001554287$ & $0.6-1.0\ \mu$m & TESS S2\\
$363$ & $2458363.87260354$ & $+0.0001581918$ & $-0.0001603849$ & $0.6-1.0\ \mu$m & TESS S2\\
$364$ & $2458364.81376481$ & $+0.0001590161$ & $-0.0001616745$ & $0.6-1.0\ \mu$m & TESS S2\\
$365$ & $2458365.75526545$ & $+0.0001551888$ & $-0.0001560597$ & $0.6-1.0\ \mu$m & TESS S2\\
$366$ & $2458366.69709569$ & $+0.0001624828$ & $-0.0001675395$ & $0.6-1.0\ \mu$m & TESS S2\\
$369$ & $2458369.52125913$ & $+0.0001696489$ & $-0.000165043$ & $0.6-1.0\ \mu$m & TESS S2\\
$370$ & $2458370.46275328$ & $+0.0001658185$ & $-0.0001689851$ & $0.6-1.0\ \mu$m & TESS S2\\
$371$ & $2458371.40401605$ & $+0.0001579167$ & $-0.0001617904$ & $0.6-1.0\ \mu$m & TESS S2\\
$372$ & $2458372.34537126$ & $+0.0001652232$ & $-0.0001639084$ & $0.6-1.0\ \mu$m & TESS S2\\
$373$ & $2458373.28731853$ & $+0.0001626426$ & $-0.0001615565$ & $0.6-1.0\ \mu$m & TESS S2\\
$374$ & $2458374.22820482$ & $+0.0001543535$ & $-0.0001656061$ & $0.6-1.0\ \mu$m & TESS S2\\
$375$ & $2458375.16979404$ & $+0.0001626522$ & $-0.0001585855$ & $0.6-1.0\ \mu$m & TESS S2\\
$376$ & $2458376.11125522$ & $+0.0001772189$ & $-0.0001689486$ & $0.6-1.0\ \mu$m & TESS S2\\
$377$ & $2458377.05267219$ & $+0.0001583615$ & $-0.0001536757$ & $0.6-1.0\ \mu$m & TESS S2\\
$378$ & $2458377.99447371$ & $+0.000163249$ & $-0.00015706$ & $0.6-1.0\ \mu$m & TESS S2\\
$379$ & $2458378.93575739$ & $+0.0001563332$ & $-0.0001569462$ & $0.6-1.0\ \mu$m & TESS S2\\
$380$ & $2458379.87716368$ & $+0.0001624613$ & $-0.0001570392$ & $0.6-1.0\ \mu$m & TESS S2\\
$381$ & $2458380.8188868$ & $+0.0001584449$ & $-0.0001654274$ & $0.6-1.0\ \mu$m & TESS S2\\
$387$ & $2458386.46725968$ & $+0.0001516065$ & $-0.0001555054$ & $0.6-1.0\ \mu$m & TESS S3\\
$388$ & $2458387.40884281$ & $+0.0001566545$ & $-0.0001593569$ & $0.6-1.0\ \mu$m & TESS S3\\
$389$ & $2458388.35020331$ & $+0.0001558977$ & $-0.000153218$ & $0.6-1.0\ \mu$m & TESS S3\\
$390$ & $2458389.29167884$ & $+0.0001504402$ & $-0.0001515723$ & $0.6-1.0\ \mu$m & TESS S3\\
$391$ & $2458390.23329699$ & $+0.000153637$ & $-0.0001603253$ & $0.6-1.0\ \mu$m & TESS S3\\
$392$ & $2458391.17449162$ & $+0.0001584431$ & $-0.0001643114$ & $0.6-1.0\ \mu$m & TESS S3\\
$393$ & $2458392.11599167$ & $+0.0001552277$ & $-0.0001539209$ & $0.6-1.0\ \mu$m & TESS S3\\
$394$ & $2458393.05747866$ & $+0.0001625986$ & $-0.0001583644$ & $0.6-1.0\ \mu$m & TESS S3\\
$395$ & $2458393.99897112$ & $+0.0001548479$ & $-0.0001582581$ & $0.6-1.0\ \mu$m & TESS S3\\
$396$ & $2458394.94021869$ & $+0.0001668753$ & $-0.000169203$ & $0.6-1.0\ \mu$m & TESS S3\\
$398$ & $2458396.82307126$ & $+0.0001512507$ & $-0.0001505658$ & $0.6-1.0\ \mu$m & TESS S3\\
$399$ & $2458397.76447778$ & $+0.000146889$ & $-0.0001459783$ & $0.6-1.0\ \mu$m & TESS S3\\
$400$ & $2458398.70654418$ & $+0.0001594393$ & $-0.0001563517$ & $0.6-1.0\ \mu$m & TESS S3\\
$401$ & $2458399.64746347$ & $+0.0001560997$ & $-0.0001559248$ & $0.6-1.0\ \mu$m & TESS S3\\
$402$ & $2458400.58901982$ & $+0.0001530252$ & $-0.0001545594$ & $0.6-1.0\ \mu$m & TESS S3\\
$403$ & $2458401.53082095$ & $+0.0001507412$ & $-0.0001534252$ & $0.6-1.0\ \mu$m & TESS S3\\
$404$ & $2458402.47208663$ & $+0.0001577229$ & $-0.0001532313$ & $0.6-1.0\ \mu$m & TESS S3\\
$405$ & $2458403.41355656$ & $+0.0001517973$ & $-0.0001556871$ & $0.6-1.0\ \mu$m & TESS S3\\
$406$ & $2458404.35498739$ & $+0.0001610387$ & $-0.0001595482$ & $0.6-1.0\ \mu$m & TESS S3\\
$407$ & $2458405.29600312$ & $+0.0001597197$ & $-0.0001557732$ & $0.6-1.0\ \mu$m & TESS S3\\
$1133$ & $2459088.79056522$ & $+0.0001600377$ & $-0.000157913$ & $0.6-1.0\ \mu$m & TESS S29\\
$1134$ & $2459089.73230357$ & $+0.00016285$ & $-0.0001653548$ & $0.6-1.0\ \mu$m & TESS S29\\
$1135$ & $2459090.67382584$ & $+0.0001615088$ & $-0.000156747$ & $0.6-1.0\ \mu$m & TESS S29\\
$1136$ & $2459091.61540082$ & $+0.0001579316$ & $-0.0001647125$ & $0.6-1.0\ \mu$m & TESS S29\\
$1137$ & $2459092.55650709$ & $+0.000163614$ & $-0.0001653631$ & $0.6-1.0\ \mu$m & TESS S29\\
$1138$ & $2459093.49800326$ & $+0.0001791626$ & $-0.000182826$ & $0.6-1.0\ \mu$m & TESS S29\\
$1139$ & $2459094.43932496$ & $+0.00016946$ & $-0.0001692725$ & $0.6-1.0\ \mu$m & TESS S29\\
$1140$ & $2459095.38106862$ & $+0.0001587269$ & $-0.0001569717$ & $0.6-1.0\ \mu$m & TESS S29\\
$1141$ & $2459096.322251$ & $+0.0001754$ & $-0.0001693464$ & $0.6-1.0\ \mu$m & TESS S29\\
$1142$ & $2459097.26367608$ & $+0.0001623146$ & $-0.000163599$ & $0.6-1.0\ \mu$m & TESS S29\\
$1143$ & $2459098.20530848$ & $+0.0001658825$ & $-0.0001709083$ & $0.6-1.0\ \mu$m & TESS S29\\
$1148$ & $2459102.91287775$ & $+0.0001981444$ & $-0.0001957247$ & $0.6-1.0\ \mu$m & TESS S29\\
$1149$ & $2459103.85470132$ & $+0.0001778397$ & $-0.0001803071$ & $0.6-1.0\ \mu$m & TESS S29\\
$1150$ & $2459104.79555654$ & $+0.0001625176$ & $-0.0001631086$ & $0.6-1.0\ \mu$m & TESS S29\\
$1151$ & $2459105.73693338$ & $+0.0001674303$ & $-0.0001728937$ & $0.6-1.0\ \mu$m & TESS S29\\
$1152$ & $2459106.67848308$ & $+0.0001915787$ & $-0.0001836274$ & $0.6-1.0\ \mu$m & TESS S29\\
$1153$ & $2459107.61998804$ & $+0.0001852554$ & $-0.0001908498$ & $0.6-1.0\ \mu$m & TESS S29\\
$1154$ & $2459108.56109682$ & $+0.0001617988$ & $-0.0001577111$ & $0.6-1.0\ \mu$m & TESS S29\\
$1155$ & $2459109.50246742$ & $+0.0001534618$ & $-0.0001552651$ & $0.6-1.0\ \mu$m & TESS S29\\
$1156$ & $2459110.44401564$ & $+0.0001615148$ & $-0.0001572489$ & $0.6-1.0\ \mu$m & TESS S29\\
$1157$ & $2459111.38546737$ & $+0.0001939066$ & $-0.0001912813$ & $0.6-1.0\ \mu$m & TESS S29\\
$1162$ & $2459116.09285056$ & $+0.0001631719$ & $-0.0001617505$ & $0.6-1.0\ \mu$m & TESS S30\\
$1163$ & $2459117.03434091$ & $+0.000159676$ & $-0.0001590436$ & $0.6-1.0\ \mu$m & TESS S30\\
$1164$ & $2459117.97587372$ & $+0.0001570867$ & $-0.0001617813$ & $0.6-1.0\ \mu$m & TESS S30\\
$1165$ & $2459118.91736845$ & $+0.0001536699$ & $-0.0001490039$ & $0.6-1.0\ \mu$m & TESS S30\\
$1166$ & $2459119.85879866$ & $+0.0001564968$ & $-0.0001585761$ & $0.6-1.0\ \mu$m & TESS S30\\
$1167$ & $2459120.80037428$ & $+0.0001553485$ & $-0.0001542201$ & $0.6-1.0\ \mu$m & TESS S30\\
$1168$ & $2459121.74156424$ & $+0.0001554497$ & $-0.0001566477$ & $0.6-1.0\ \mu$m & TESS S30\\
$1169$ & $2459122.68320566$ & $+0.0001682892$ & $-0.0001619783$ & $0.6-1.0\ \mu$m & TESS S30\\
$1170$ & $2459123.62416204$ & $+0.0001632229$ & $-0.0001626586$ & $0.6-1.0\ \mu$m & TESS S30\\
$1171$ & $2459124.56604724$ & $+0.0001482321$ & $-0.0001548794$ & $0.6-1.0\ \mu$m & TESS S30\\
$1172$ & $2459125.50752307$ & $+0.0001542799$ & $-0.0001514558$ & $0.6-1.0\ \mu$m & TESS S30\\
$1173$ & $2459126.44901138$ & $+0.0001535523$ & $-0.0001585222$ & $0.6-1.0\ \mu$m & TESS S30\\
$1178$ & $2459131.15613401$ & $+0.0001649443$ & $-0.000162974$ & $0.6-1.0\ \mu$m & TESS S30\\
$1179$ & $2459132.09748432$ & $+0.000152724$ & $-0.0001505922$ & $0.6-1.0\ \mu$m & TESS S30\\
$1180$ & $2459133.03891429$ & $+0.000160387$ & $-0.0001600435$ & $0.6-1.0\ \mu$m & TESS S30\\
$1181$ & $2459133.98016695$ & $+0.0001660438$ & $-0.0001685588$ & $0.6-1.0\ \mu$m & TESS S30\\
$1182$ & $2459134.92223374$ & $+0.0001939714$ & $-0.000202803$ & $0.6-1.0\ \mu$m & TESS S30\\
$1183$ & $2459135.86338806$ & $+0.000181922$ & $-0.000183373$ & $0.6-1.0\ \mu$m & TESS S30\\
$1184$ & $2459136.80530221$ & $+0.0001828732$ & $-0.0001829882$ & $0.6-1.0\ \mu$m & TESS S30\\
$1185$ & $2459137.74616908$ & $+0.0001570621$ & $-0.000162082$ & $0.6-1.0\ \mu$m & TESS S30\\
$1186$ & $2459138.6877949$ & $+0.0001539593$ & $-0.0001510713$ & $0.6-1.0\ \mu$m & TESS S30\\
$1187$ & $2459139.62954128$ & $+0.0001504001$ & $-0.000146267$ & $0.6-1.0\ \mu$m & TESS S30\\
$1188$ & $2459140.57066808$ & $+0.0001572311$ & $-0.000159404$ & $0.6-1.0\ \mu$m & TESS S30\\
$1189$ & $2459141.51195794$ & $+0.0001686196$ & $-0.0001678828$ & $0.6-1.0\ \mu$m & TESS S30\\
\end{longtable}

\begin{longtable}{r l l l l l}
\caption{WASP-19b transit times with epochs calculated from reference time $T_\mathrm{0}=2456021.70389628$.}
\label{w19_times} \\     % is used to refer this table in the text

\hline\hline             
Epoch & $T_\mathrm{mid}$ (BJD) & \multicolumn{2}{c}{Residuals} & Bandpass & Light Curve Reference \\    % table heading 
\hline 
\endfirsthead  

\multicolumn{6}{c}
{{\tablename\ \thetable{} -- continued}} \\
\hline Epoch & $T_\mathrm{mid}$ (BJD) & \multicolumn{2}{c}{Residuals} & Bandpass & Light Curve Reference \\  \hline 
\endhead

\hline \multicolumn{6}{r}{{Continues on the next page}} \\ \hline
\endfoot

\hline
\endlastfoot

$-1580$ & $2454775.3372$ & $+0.0015$ & $-0.0015$ & \textit{z} & \cite{hebb2009wasp} \\
$-1578$ & $2454776.91566$ & $+0.00019$ & $-0.00019$ & \textit{H} & \cite{anderson2010h} \\
$-1527$ & $2454817.14633$ & $+0.00021$ & $-0.00021$ & \textit{H} & \cite{lendl2013photometric} \\
$-1042$ & $2455199.73343$ & $+0.00083$ & $-0.00083$ & Clear & Tifner F. (TRESCA)\\ 
$-976$ & $2455251.79657$ & $+0.00014$ & $-0.00014$ & \textit{r-Gunn} & \cite{tregloan2013transits} \\
$-975$ & $2455252.58544$ & $+0.0001$ & $-0.0001$ & \textit{r-Gunn} & \cite{tregloan2013transits} \\
$-971$ & $2455255.74077$ & $+0.00012$ & $-0.00012$ & \textit{r-Gunn} & \cite{tregloan2013transits} \\
$-966$ & $2455259.68448$ & $+0.00033$ & $-0.00033$ & Clear & Coloque J. (TRESCA)\\
$-948$ & $2455273.88282$ & $+0.00062$ & $-0.00062$ & Clear & Evans P. (TRESCA)\\
$-916$ & $2455299.12768$ & $+0.00055$ & $-0.00055$ & Clear & Curtis I. (TRESCA)\\
$-871$ & $2455334.6254$ & $+0.00021$ & $-0.00021$ & \textit{R} & \cite{mancini2013physical} \\
$-866$ & $2455338.56927$ & $+0.00023$ & $-0.00023$ & \textit{I+z'} & \cite{lendl2013photometric}\\
$-847$ & $2455353.55659$ & $+0.00024$ & $-0.00024$ & \textit{R} & \cite{mancini2013physical}\\
$-834$ & $2455363.81131$ & $+0.00041$ & $-0.00041$ & \textit{BB} & Milne G. (TRESCA)\\
$-828$ & $2455368.54285$ & $+0.00212$ & $-0.00212$ & \textit{R} & \cite{mancini2013physical}\\
$-611$ & $2455539.72327$ & $+0.0003$ & $-0.0003$ & \textit{I-Cousins} & \cite{lendl2013photometric}\\
$-573$ & $2455569.69826$ & $+0.00036$ & $-0.00036$ & \textit{I+z'} & \cite{lendl2013photometric} \\
$-556$ & $2455583.10979$ & $+0.00089$ & $-0.00089$ & Clear & Curtis I. (TRESCA)\\
$-554$ & $2455584.68693$ & $+0.00024$ & $-0.00024$ & \textit{r'} & \cite{lendl2013photometric}\\
$-554$ & $2455584.68684$ & $+0.00019$ & $-0.00019$ & \textit{I+z'} & \cite{lendl2013photometric}\\
$-542$ & $2455594.15188$ & $+0.00168$ & $-0.00168$ & \textit{R} & \cite{mancini2013physical}\\
$-533$ & $2455601.25164$ & $+0.00071$ & $-0.00071$ & \textit{R} & \cite{mancini2013physical}\\
$-531$ & $2455602.83138$ & $+0.00046$ & $-0.00046$ & \textit{I+z'} & \cite{lendl2013photometric}\\
$-528$ & $2455605.19414$ & $+0.0018$ & $-0.0018$ & \textit{R} & \cite{mancini2013physical}\\
$-526$ & $2455606.77464$ & $+0.00022$ & $-0.00022$ & \textit{z'} & \cite{lendl2013photometric}\\
$-525$ & $2455607.56241$ & $+0.00033$ & $-0.00033$ & \textit{I+z'} & \cite{lendl2013photometric}\\
$-506$ & $2455622.55057$ & $+0.00026$ & $-0.00026$ & \textit{I+z'} & \cite{lendl2013photometric}\\
$-504$ & $2455624.12787$ & $+0.00142$ & $-0.00142$ & \textit{R} & \cite{mancini2013physical}\\
$-493$ & $2455632.80612$ & $+0.00025$ & $-0.00025$ & \textit{r'} & \cite{lendl2013photometric}\\
$-464$ & $2455655.68222$ & $+0.00045$ & $-0.00045$ & \textit{I+z'} & \cite{lendl2013photometric}\\
$-445$ & $2455670.66976$ & $+0.00064$ & $-0.00064$ & \textit{I+z'} & \cite{lendl2013photometric}\\
$-436$ & $2455677.77038$ & $+0.00195$ & $-0.00195$ & \textit{R} & \cite{mancini2013physical}\\
$-422$ & $2455688.81201$ & $+0.00333$ & $-0.00333$ & \textit{R} & \cite{mancini2013physical}\\
$-421$ & $2455689.60276$ & $+0.0003$ & $-0.0003$ & \textit{R} & \cite{mancini2013physical}\\
$-417$ & $2455692.75674$ & $+0.00255$ & $-0.00255$ & \textit{R} & \cite{mancini2013physical}\\
$-416$ & $2455693.54639$ & $+0.00013$ & $-0.00013$ & \textit{R} & \cite{mancini2013physical}\\
$-403$ & $2455703.79933$ & $+0.00411$ & $-0.00411$ & \textit{R} & \cite{mancini2013physical}\\
$-402$ & $2455704.59078$ & $+0.00034$ & $-0.00034$ & \textit{R} & \cite{mancini2013physical}\\
$-397$ & $2455708.53495$ & $+0.00015$ & $-0.00015$ & \textit{R} & \cite{mancini2013physical}\\
$-171$ & $2455886.81234$ & $+0.00208$ & $-0.00208$ & \textit{R} & \cite{mancini2013physical}\\
$-159$ & $2455896.27611$ & $+0.0021$ & $-0.0021$ & \textit{R} & \cite{mancini2013physical}\\
$-135$ & $2455915.2098$ & $+0.00065$ & $-0.00065$ & \textit{R} & \cite{mancini2013physical}\\
$-130$ & $2455919.15485$ & $+0.00103$ & $-0.00103$ & \textit{R} & \cite{mancini2013physical}\\
$-126$ & $2455922.30966$ & $+0.00555$ & $-0.00555$ & \textit{R} & \cite{mancini2013physical}\\
$-28$ & $2455999.616301$ & $+0.00007$ & $-0.00007$ & \textit{HK} & \cite{bean2013ground}\\
$0$ & $2456021.70374$ & $+0.000085$ & $-0.000085$ & \textit{HK} & \cite{bean2013ground}\\
$10$ & $2456029.5925$ & $+0.00035$ & $-0.00035$ & \textit{r'} & \cite{lendl2013photometric}\\
$15$ & $2456033.53645$ & $+0.00014$ & $-0.00014$ & \textit{g'} & \cite{mancini2013physical}\\
$15$ & $2456033.53651$ & $+0.00007$ & $-0.00007$ & \textit{r'} & \cite{mancini2013physical}\\
$15$ & $2456033.53643$ & $+0.00007$ & $-0.00007$ & \textit{i'} & \cite{mancini2013physical}\\
$15$ & $2456033.53652$ & $+0.00009$ & $-0.00009$ & \textit{z'} & \cite{mancini2013physical}\\
$53$ & $2456063.51174$ & $+0.0003$ & $-0.0003$ & \textit{I+z'} & \cite{lendl2013photometric} \\
$397$ & $2456334.87208$ & $+0.00053$ & $-0.00053$ & Clear & Evans P. (TRESCA)\\
$910$ & $2456739.547178$ & $+0.000046$ & $-0.000046$ & White light & \cite{espinoza2019access}\\
$957$ & $2456776.622511$ & $+0.000066$ & $-0.000066$ & White light & \cite{espinoza2019access}\\
$1464$ & $2457176.563965$ & $+0.000094$ & $-0.000094$ & White light & \cite{espinoza2019access}\\
$2250$ & $2457796.591441$ & $+0.000061$ & $-0.000061$ & White light & \cite{espinoza2019access}\\
$2316$ & $2457848.654791$ & $+0.000068$ & $-0.000068$ & White light & \cite{espinoza2019access}\\
$2326$ & $2457856.543042$ & $+0.000111$ & $-0.000111$ & White light & \cite{espinoza2019access}\\
$952$ & $2456772.67789$ & $+0.00052$ & $-0.00052$ & \textit{I+z'} & \cite{patra2020continuing}\\
$1933$ & $2457546.53025$ & $+0.00038$ & $-0.00038$ & \textit{I+z'} & \cite{patra2020continuing}\\
$3244$ & $2458580.69724$ & $+0.0004$ & $-0.0004$ & \textit{I+z'} & \cite{patra2020continuing}\\
$3198$ & $2458544.41098726$ & $+0.0005096953$ & $-0.0005161792$ & $0.6-1.0\ \mu$m & TESS S9\\
$3199$ & $2458545.19909857$ & $+0.0004676348$ & $-0.0004776159$ & $0.6-1.0\ \mu$m & TESS S9\\
$3200$ & $2458545.98899546$ & $+0.000459334$ & $-0.0004546766$ & $0.6-1.0\ \mu$m & TESS S9\\
$3201$ & $2458546.77724425$ & $+0.0004526817$ & $-0.0004699151$ & $0.6-1.0\ \mu$m & TESS S9\\
$3202$ & $2458547.56694414$ & $+0.0004957815$ & $-0.0005057116$ & $0.6-1.0\ \mu$m & TESS S9\\
$3203$ & $2458548.35481158$ & $+0.0004678426$ & $-0.0004630053$ & $0.6-1.0\ \mu$m & TESS S9\\
$3204$ & $2458549.14425338$ & $+0.0004675586$ & $-0.0004538751$ & $0.6-1.0\ \mu$m & TESS S9\\
$3205$ & $2458549.93228826$ & $+0.0004875952$ & $-0.0004723997$ & $0.6-1.0\ \mu$m & TESS S9\\
$3206$ & $2458550.72132794$ & $+0.0004917443$ & $-0.0004904705$ & $0.6-1.0\ \mu$m & TESS S9\\
$3207$ & $2458551.51006035$ & $+0.0005079718$ & $-0.000499857$ & $0.6-1.0\ \mu$m & TESS S9\\
$3208$ & $2458552.29893896$ & $+0.000491923$ & $-0.0004829233$ & $0.6-1.0\ \mu$m & TESS S9\\
$3209$ & $2458553.08800457$ & $+0.000490958$ & $-0.0004977896$ & $0.6-1.0\ \mu$m & TESS S9\\
$3210$ & $2458553.87691279$ & $+0.000444077$ & $-0.0004363738$ & $0.6-1.0\ \mu$m & TESS S9\\
$3211$ & $2458554.66571083$ & $+0.0004551007$ & $-0.0004432428$ & $0.6-1.0\ \mu$m & TESS S9\\
$3212$ & $2458555.45433949$ & $+0.0004878705$ & $-0.0004864196$ & $0.6-1.0\ \mu$m & TESS S9\\
$3215$ & $2458557.82151997$ & $+0.0005212972$ & $-0.0005195209$ & $0.6-1.0\ \mu$m & TESS S9\\
$3216$ & $2458558.609454$ & $+0.0004931412$ & $-0.000493327$ & $0.6-1.0\ \mu$m & TESS S9\\
$3217$ & $2458559.39915513$ & $+0.0004648924$ & $-0.000459963$ & $0.6-1.0\ \mu$m & TESS S9\\
$3218$ & $2458560.18683688$ & $+0.0004597507$ & $-0.0004630294$ & $0.6-1.0\ \mu$m & TESS S9\\
$3219$ & $2458560.97725509$ & $+0.0004797756$ & $-0.0004820024$ & $0.6-1.0\ \mu$m & TESS S9\\
$3220$ & $2458561.76548315$ & $+0.0004690496$ & $-0.0004596392$ & $0.6-1.0\ \mu$m & TESS S9\\
$3221$ & $2458562.55375174$ & $+0.0004630492$ & $-0.0004569487$ & $0.6-1.0\ \mu$m & TESS S9\\
$3222$ & $2458563.34272117$ & $+0.000474578$ & $-0.0004586996$ & $0.6-1.0\ \mu$m & TESS S9\\
$3223$ & $2458564.13193688$ & $+0.0004401143$ & $-0.000421447$ & $0.6-1.0\ \mu$m & TESS S9\\
$3224$ & $2458564.92115668$ & $+0.0004787526$ & $-0.0004844793$ & $0.6-1.0\ \mu$m & TESS S9\\
$3225$ & $2458565.70935215$ & $+0.0004918357$ & $-0.000486038$ & $0.6-1.0\ \mu$m & TESS S9\\
$3226$ & $2458566.49875766$ & $+0.000457878$ & $-0.0004531755$ & $0.6-1.0\ \mu$m & TESS S9\\
$3227$ & $2458567.28771693$ & $+0.000479046$ & $-0.0004815575$ & $0.6-1.0\ \mu$m & TESS S9\\
$3228$ & $2458568.07636429$ & $+0.0004715116$ & $-0.000471451$ & $0.6-1.0\ \mu$m & TESS S9\\
$4134$ & $2459282.76422855$ & $+0.000445984$ & $-0.0004382528$ & $0.6-1.0\ \mu$m & TESS S36\\
$4135$ & $2459283.55349494$ & $+0.0004680825$ & $-0.0004832548$ & $0.6-1.0\ \mu$m & TESS S36\\
$4136$ & $2459284.34243336$ & $+0.0005023551$ & $-0.000509186$ & $0.6-1.0\ \mu$m & TESS S36\\
$4137$ & $2459285.13027096$ & $+0.0004852793$ & $-0.0004831051$ & $0.6-1.0\ \mu$m & TESS S36\\
$4138$ & $2459285.91977053$ & $+0.0004735948$ & $-0.0004618111$ & $0.6-1.0\ \mu$m & TESS S36\\
$4139$ & $2459286.70892877$ & $+0.0005314286$ & $-0.0005276405$ & $0.6-1.0\ \mu$m & TESS S36\\
$4140$ & $2459287.49727343$ & $+0.0004814359$ & $-0.000498216$ & $0.6-1.0\ \mu$m & TESS S36\\
$4141$ & $2459288.28588867$ & $+0.0004538875$ & $-0.0004737029$ & $0.6-1.0\ \mu$m & TESS S36\\
$4142$ & $2459289.07496964$ & $+0.0005078695$ & $-0.000492987$ & $0.6-1.0\ \mu$m & TESS S36\\
$4143$ & $2459289.86351056$ & $+0.0004664274$ & $-0.0004644744$ & $0.6-1.0\ \mu$m & TESS S36\\
$4144$ & $2459290.65262719$ & $+0.0005161349$ & $-0.0005024358$ & $0.6-1.0\ \mu$m & TESS S36\\
$4145$ & $2459291.44091157$ & $+0.0005141344$ & $-0.0005178709$ & $0.6-1.0\ \mu$m & TESS S36\\
$4146$ & $2459292.23014503$ & $+0.0004648569$ & $-0.0004685227$ & $0.6-1.0\ \mu$m & TESS S36\\
$4150$ & $2459295.38506316$ & $+0.0005516386$ & $-0.0005686305$ & $0.6-1.0\ \mu$m & TESS S36\\
$4151$ & $2459296.17521974$ & $+0.0004610964$ & $-0.0004679605$ & $0.6-1.0\ \mu$m & TESS S36\\
$4152$ & $2459296.9630376$ & $+0.0004828338$ & $-0.0004728424$ & $0.6-1.0\ \mu$m & TESS S36\\
$4153$ & $2459297.75137559$ & $+0.0005018343$ & $-0.000503573$ & $0.6-1.0\ \mu$m & TESS S36\\
$4154$ & $2459298.53984262$ & $+0.0004852697$ & $-0.0004830355$ & $0.6-1.0\ \mu$m & TESS S36\\
$4155$ & $2459299.32941899$ & $+0.0004804281$ & $-0.0004758144$ & $0.6-1.0\ \mu$m & TESS S36\\
$4156$ & $2459300.11844447$ & $+0.0004893952$ & $-0.0004878233$ & $0.6-1.0\ \mu$m & TESS S36\\
$4157$ & $2459300.90763443$ & $+0.0004986904$ & $-0.0005087346$ & $0.6-1.0\ \mu$m & TESS S36\\
$4158$ & $2459301.69651184$ & $+0.0004741549$ & $-0.0004809383$ & $0.6-1.0\ \mu$m & TESS S36\\
$4159$ & $2459302.48452828$ & $+0.0004956919$ & $-0.0004803605$ & $0.6-1.0\ \mu$m & TESS S36\\
$4160$ & $2459303.27444271$ & $+0.0005392028$ & $-0.0005249785$ & $0.6-1.0\ \mu$m & TESS S36\\
$4161$ & $2459304.06362531$ & $+0.0004885187$ & $-0.0004994962$ & $0.6-1.0\ \mu$m & TESS S36\\
$4162$ & $2459304.85196345$ & $+0.000436011$ & $-0.0004488881$ & $0.6-1.0\ \mu$m & TESS S36\\
$4163$ & $2459305.64011759$ & $+0.000536406$ & $-0.0005166219$ & $0.6-1.0\ \mu$m & TESS S36\\

\end{longtable}

\end{appendix}

\end{document}